\documentclass[a4paper,11pt]{article}
\pdfoutput=1
\usepackage{jheppub}

\usepackage[T1]{fontenc}
\usepackage{amsmath}
\usepackage{amssymb, amsthm,graphicx}
\newcommand{\mathsym}[1]{{}}
\newcommand{\unicode}[1]{{}}

\newcommand{\be}{\begin{equation}}
\newcommand{\ee}{\end{equation}}

\def\beq{\begin{eqnarray}}\def\eeq{\end{eqnarray}}
\def\be{\begin{equation}}\def\ee{\end{equation}}

\preprint{TIFR/TH/15-17}

\title{ The Shear Viscosity in Anisotropic Phases}

\author{ Sachin Jain$^{1}$, Rickmoy Samanta$^{2}$ and Sandip P. Trivedi$^{2}$\\
\it $^{1}$ Department of Physics,
\it Cornell University, Ithaca, New York 14853, USA.\\
\it $^{2}$ Department of Theoretical Physics,
\it Tata Institute of Fundamental Research,\\ \it Colaba, Mumbai, 400005, India}

\emailAdd{sj339@cornell.edu} 
\emailAdd{rickmoy@theory.tifr.res.in} 
\emailAdd{trivedi.sp@gmail.com} 

\vspace{0.5cm}

\abstract{We construct anisotropic black brane solutions and analyse the behaviour of some of their metric perturbations.  These  solutions correspond to field theory duals in which 
  rotational symmetry  is broken due  an externally applied, spatially constant, force. We find, in several examples,  that when the anisotropy is sufficiently big compared to the temperature, some components of the viscosity tensor can become very small
  in units of  the entropy density, parametrically violating the KSS bound. We obtain an expression  relating  these  components of the viscosity, in units of the entropy density, to a ratio of metric components at the horizon of the black brane. This relation is  generally valid, as long as the forcing function is translationally invariant,  and it directly connects  the parametric violation of the bound to the anisotropy in the metric at the horizon.  Our results  suggest the possibility that such small components of the viscosity tensor  might also arise in 
  anisotropic strongly coupled fluids found in nature. 
  }

\begin{document}
\maketitle
\tableofcontents

\section{Introduction}
\label{intro}

The AdS/CFT correspondence has emerged as an important tool in  the analysis of strongly coupled systems, especially for the  study  of transport properties of such systems.   Neither analytical nor numerical methods are convenient for  calculating these properties  on the field theory  side since they require an understanding of the real time response at finite temperature.  In contrast, they can be calculated with relative ease on the gravity side, often by solving simple linear equations.   
An important insight  which has come out of these studies pertains to  the behaviour of the viscosity. It was found in KSS \cite{Policastro:2001yc,Kovtun:2003wp,Kovtun:2004de}, that for systems having a gravity description that can be well approximated by classical Einstein gravity, the ratio of the shear viscosity, $\eta$, to the entropy density, $s$, takes the universal value
 \be
 \label{kss}
 {\eta \over s}={1\over 4 \pi}.
 \ee
This is a small value, compared to weak coupling where the ratio diverges. It was also initially suggested that this value is  a bound,  and the  ratio  can never become smaller. 
We now know that this is not true \cite{Kats:2007mq,Buchel:2008vz,Sinha:2009ev,Cremonini:2011iq}, see also \cite{Basu:2011tt,Bhattacharyya:2014wfa}, but in all controlled counter-examples the bound is violated at best  by a numerical factor, and not in a parametric manner. 
Attempts to produce bigger violations lead to physically unacceptable situations, e.g., to causality violations, for example, see \cite{Brigante:2007nu,Brigante:2008gz}. 
However, there is some discussion  of  a violation  of the bound in metastable states, see \cite{Cohen:2007qr}. Also, see  \cite{Buchel:2010wf} for a discussion of violations in a superfluid phase described by higher derivative gravity.

The behaviour of the viscosity discussed  above refers to isotropic and homogeneous phases, which on the gravity side at finite temperature are described by the Schwarzschild black brane geometry. More recently, gravitational backgrounds which correspond to anisotropic phases in field theory have also been studied in \cite{Landsteiner:2007bd,Azeyanagi:2009pr,Natsuume:2010ky,Erdmenger:2010xm,Basu:2011tt,Erdmenger:2011tj,Mateos:2011ix,Mateos:2011tv,Cheng:2014sxa} and the behaviour of the  viscosity in  some  of these anisotropic phases has also been analysed, see \cite{Rebhan:2011vd, Polchinski:2012nh} and \cite{Giataganas:2012zy,Iizuka:2012wt,Mamo:2012sy,Jain:2014vka,Critelli:2014kra,Ge:2014aza}. The viscosity in the anisotropic case is  a tensor, which in the most general case, with no rotational invariance, has 21 independent components (when the field theory lives in  $3+1$ dimensions). 
In \cite{Rebhan:2011vd,Polchinski:2012nh,Jain:2014vka}, where some simple cases were considered, it was found that some components of the viscosity tensor can become much smaller, parametrically violating the bound in eq.(\ref{kss}). 
For example, in \cite{Jain:2014vka}, a gravitational solution was considered where  the rotational invariance of the three space dimensions in which the field theory lives was broken from $SO(3)$ to $SO(2)$, due to a linearly varying dilaton .
In the solution, the  dilaton varies  along the $z$ direction and rotational invariance in the remaining $x,y$,  spatial directions was left unbroken. 
 The component of the viscosity, called $\eta_{||}$ in \cite{Jain:2014vka}, which measures the shear force in the $x-y$ plane,  was still  found to satisfy the relation, eq.(\ref{kss}).
However,  other  components of the viscosity did not satisfy it. In particular, it was found that a component called $\eta_\perp$, which measures the shear force in the $x-z$ or $y-z$ plane, could become much smaller, going like 

 \be
 \label{exan}
 {\eta_{\perp}\over s} = {8 \pi \over 3} {T^2\over \rho^2},
 \ee
where $T$ is the temperature and $\rho$ is the anisotropy parameter. The result, eq.(\ref{exan}) is valid in the extremely anisotropic limit, when $T\ll \rho$. A detailed study was also carried out in \cite{Jain:2014vka} of  this extreme anisotropic regime  and no instabilities were found to be present. 

In this paper we study many other examples where anisotropic phases arise and show that in all of them components of the viscosity can become parametrically small, in units of the entropy density, when the anisotropy becomes sufficiently large compared to the temperature. Depending on the example, the factor of $T^2$ in eq.(\ref{exan}) can be replaced by some other positive power of $T$. 

A common feature of all our examples is that the breaking of anisotropy is due to an externally applied force which is translationally invariant. For example,  the linearly varying dilaton considered in \cite{Jain:2014vka}, and also in section (\ref{rev1dil})  gives rise to such a spatially constant forcing function. This follows from the fact that the boundary theory stress tensor is no longer conserved in the presence of the dilaton and instead satisfies the equation 
\be
\label{repl}
\partial_\mu <T^{\mu\nu}>=<\hat{O}> \partial^\nu \phi,
\ee
where $\hat{O}$ is the operator dual to the dilaton, see eq.(6.9) of \cite{Jain:2014vka}. Similarly, we consider linearly varying axions in section \ref{oneax} and \ref{twoax}, and a constant magnetic field in section \ref{mag}.

Another common feature in our examples is that some residual Lorentz symmetry survives, at zero temperature, after incorporating the breaking of rotational invariance.  
Fluid mechanics then corresponds to the dynamics of the goldstone modes associated with the boost symmetries of this Lorentz group which are broken at finite temperature.  

In the second half of  this paper we give an argument, based on a Kaluza Klein decomposition of modes, which shows quite generally that in all situations sharing these features, in particular where the forcing function does not break translational invariance, appropriate components of the viscosity tensor become parametrically small. These components correspond to  perturbations of the metric which carry spin $1$ with respect to the surviving Lorentz symmetry. Let 
$z$ be a spatial direction in the boundary theory along which there is anisotropy and $x$ be a spatial direction  along which the boost symmetry is left unbroken, 
then we show that the viscosity component $\eta_{xz}$, which couples to the $h_{xz}$ component of the metric perturbation, satisfies the relation,
\be\label{mresaa}{\eta_{xz}\over s} = {1\over 4\pi} {{ g}_{xx}\over { g}_{zz}}~\Big{|}_{u=u_{h}}, \ee
where ${ g}_{xx}|_{u=u_{h}}, {g}_{zz}|_{u=u_{h}}$ refer to the components of the background metric at the horizon. 
Eq.(\ref{mresaa}) is one of the main results of the paper.  It also agrees with the behaviour seen in all the explicit examples we consider. This result was first derived for an anisotropic axion-dilaton-gravity system in \cite{Rebhan:2011vd}.

In the isotropic case the ratio ${{ g}_{xx}\over { g}_{zz}}~\Big{|}_{u=u_{h}}$ is unity and we see that the KSS result  is obtained. 
However, in anisotropic cases this ratio can become very different from unity and in fact much smaller, leading to the parametric violation of the bound, eq.(\ref{kss}). 

Note that the result, eq.(\ref{mresaa}), is true for conformally invariant systems, as well as systems with a mass gap, when subjected to a constant driving force. 
Examples of massive systems include,  for example,   gravitational duals of confining gauge theories, \cite{Witten:1998zw} and \cite{Klebanov:2000hb}. For these cases the temperature should be bigger  than the confining scale so that the gravity dual is described by a black brane.  Also, for  some components of the viscosity  to become  significantly smaller than the bound, the anisotropy must be bigger than the temperature.

\begin{figure}
\begin{center}
\includegraphics[width=0.8\textwidth]{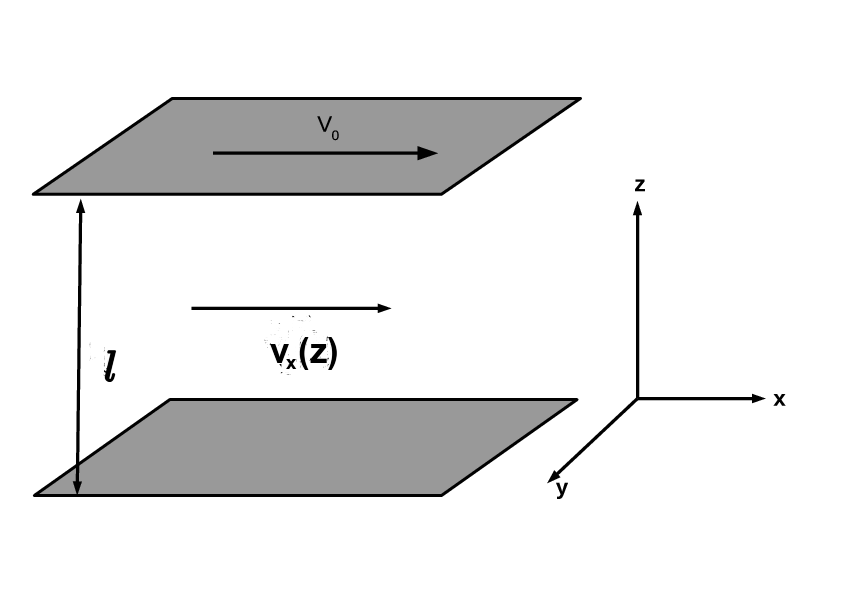}
\caption{Picture showing flow of fluid enclosed between two parallel plates separated along the $z$-direction.}
\label{plt}
\end{center}
\end{figure}

Physically  a component like $\eta_{xz}$ measures the resistance to shear. For example, if the fluid is enclosed between two parallel plates which are separated along the $z$ direction and moving with a relative velocity $ v_x$ along the $x$ direction in a non-relativistic
fashion, they will experience a friction force due to the fluid, proportional to $\eta_{xz} \partial_zv_x$.  See Fig (\ref{plt}) and the more extensive discussion in section 6 of \cite{Jain:2014vka}. Thus the parametrically small values obtained here correspond to a very small resistance to shear in anisotropic systems.

Our results which are quite general, open up the exciting possibility that in nature too, strongly coupled anisotropic systems may  have a very small value for components of the viscosity.  It would be very exciting if this behaviour can be probed in experimental situations, realised perhaps in cold atom systems, or in the context of QCD. 

%Ordinarily, QCD at finite temperature is described by a  homogeneous and isotropic phase for which  the calculations discussed here are not relevant.
%This is true even when we consider situations which come about  due  to anisotropic initial conditions, as might arise in heavy ion collisions. The behaviour of the QCD fluid   in these situations is still  governed by  rotationally invariant Navier Stokes equations with appropriate viscosity coefficients.  However, this could change if a sufficiently big magnetic field   is turned on breaking rotational invariance \footnote{A magnetic field of order $10^{16}$ Tesla or so is needed in order to contribute an energy density comparable to the QCD scale $\sim 200$ Mev.}.  The resulting  equilibrium phase  could then be highly anisotropic and our results hint that  suitable components of the viscosity might  perhaps then become small. 
%This might perhaps be relevant in the interior of highly magnetised neutron stars, see \cite{Endrodi:2014lja}, \cite{Bocquet:1995je} and \cite{Harding:2006qn}.  It has  also been suggested that such intense magnetic fields might  actually  arise in heavy ion collisions, see \cite{Skokov:2009qp}, since the ions which collide are electrically charged and highly relativistic, although in this case the transitory nature of these fields must also then be taken into account.    

This paper is structured as follows. In section \ref{rev1dil} we  review the earlier discussion of a system with one  linearly varying dilaton. Some  general aspects involved in  the calculation of viscosity are discussed in section \ref{viscomp}. Several examples of anisotropic systems realised in gravity are then discussed, including the case with two dilatons in section \ref{gravitysolscalar}, a magnetic field in section \ref{mag}, and axions and dilatons, section \ref{oneax} and section \ref{twoax}. The general argument based on a Kaluza Klein truncation is given in section \ref{kkan}. 
We end with conclusions in section \ref{cncl}. The appendices \ref{Numerics}, \ref{bdb} and \ref{b2hz} contain additional important details.

\section{Brief Review of The System With One Dilaton}
\label{rev1dil}
Here we briefly summarise  some of the key results in \cite{Jain:2014vka} which considered a linearly varying dilaton 
$\phi=\rho\, z$
in asymptotically $AdS_5$, for a theory with action 
\be \label{action}
S_{bulk}=\frac{1}{2\kappa^2}\int d^{5}x\sqrt{-g}~\left(R+12\Lambda-\frac{1}{2}(\partial\phi)^2\right).
\ee
Here $2\kappa^2=16\pi G$ is the gravitational coupling and $G$ is the Newton's Constant in 5-dimensions.
At zero temperature the near horizon solution was found to be $AdS_4 \times R$,
\begin{align}  \label{metans1}
ds^2&=-\frac{4}{3}u^2dt^2+\frac{du^2}{\frac{4}{3}u^2}+{4\over 3}~ u^2(dx^2+dy^2)+ \frac{\rho^2}{8}dz^2.
\end{align}
The  radius of $AdS_4$,  $R_4^2=3/4 $, in units  where  $\Lambda=1$. 
 We see   in eq.(\ref{metans1})  that the metric component $g_{zz}$ becomes constant due to  the extra stress energy provided by the linearly varying  dilaton. 
The $AdS_4\times R$ solution is in fact an exact solution to the equations of motion.

  At small temperature, $T\ll \rho$, the geometry is that of a Schwarzschild black brane in $AdS_{4}\times R$. The viscosity is related, using linear response, to the retarded two point function of components of the stress tensor, and the latter, using Ads/CFT, can be calculated from the behaviour of appropriate  metric perturbations in the bulk. The answer for $\eta_{xy}$,  which is denoted as $\eta_{||}$ and for $\eta_{xz}, \eta_{yz}$, which are equal and denoted as $\eta_{\perp}$, is given  in eq.(\ref{etaprll1})
and eq.(\ref{etaperpint}) below:
\begin{equation}
 \frac{\eta_{\parallel}}{s}=\frac{1}{4\pi}\label{etaprll1},
\end{equation}
\be
\label{etaperpint}
\frac{\eta_{\bot}}{s} =\frac{8\pi T^2}{3\rho^2},
\ee 
with $s$ being the entropy density. 

 We see that $\eta_\perp$ in units of  the entropy density becomes parametrically small in the limit of high anisotropy. The fluid mechanics in this high anisotropy limit was also  systematically  set up  in \cite{Jain:2014vka} and it was shown that, as expected, this small viscosity component results in a very small  shear force on two suitably oriented  parallel plates which are moving with a relative velocity and enclose the fluid. 

\section{More Details On The Calculation Of Viscosity}
\label{viscomp}
Before proceeding it is worth giving some more details on the calculation of the viscosity for the one dilaton system above. These features, as we will see, will be shared by all the examples we consider subsequently in this paper. The analysis that follows
will also reveal   the central reason for why the viscosity in units of the entropy density can become so small in anisotropic systems. 

With anisotropy, the viscosity is a tensor, $\eta_{ijkl}$,  in general with $21$ components. Using the Kubo formula these can be related to the two point function of the stress energy tensor as follows , 
\begin{equation}
\label{kuboa}
\eta_{ij,kl} = -\lim_{\omega\rightarrow 0}\frac{1}{\omega}\, \bold{Im} \big[ G^{R}_{ij,kl}(\omega)\big],
\end{equation}
where
\begin{equation}
\label{defGR}
 G^{R}_{ij,kl}(\omega,0) = \int dt ~ d{\bf x}\, e^{i \omega t}\, \theta(t)\, \langle[T_{ij}(t,{\bf x}),T_{kl}(0,0)]\rangle,
\end{equation}
 and  $\bold{Im}$ denotes the imaginary part of the retarded Green's function. \\
From the AdS/CFT correspondence the two point function of $T_{ij}$ can be calculated in terms of the behaviour of metric perturbations, and in this way the viscosity can be obtained. 

In the one dilaton system considered in section \ref{rev1dil} , the solution has an $SO(2)$ rotational invariance in the $x-y$ plane, as is evident from the metric (\ref {metans1}). For simplicity we denote the $\eta_{xz,xz}$ component as $\eta_{xz}$, and $\eta_{yz,yz}$ as $\eta_{yz}$ etc. Due to the $SO(2)$ invariance we get that $\eta_{xz}=\eta_{yz}\equiv \eta_\perp$. 
These components are related to the behaviour of the $h_{xz}, h_{yz}$ components of metric perturbations, which carry spin $1$ with respect to $SO(2)$ symmetry.  

We now proceed to introduce the $h_{xz}$ perturbation in the metric as follows
\begin{equation}
\label{defperta}
\begin{split}
  ds^2=-g_{tt}(u) dt^2+g_{uu}(u)du^2+ &g_{xx}(u)dx^2 + g_{yy}dy^2+g_{zz}(u)dz^2 \\ &+2e^{-i\omega t}Z(u)g_{xx}(u)dx~ dz,
 \end{split}
\end{equation}
where $Z(u)$ is the required perturbation of interest.  
We can show that the other modes decouple from 
$Z(u)$  and hence we can consistently set them to zero. Here we follow closely \cite{Iqbal:2008by}.

One finds that the mode $Z(u)$ obeys an equation of the form
\begin{equation}\label{eqnscal1}
 \partial_{u}\left(\sqrt{-g}P(u)g^{uu}\partial_{u}Z(u)\right)-\omega^2 N(u)g^{tt} Z(u)=0,
\end{equation} 

The  functions $P(u), N(u)$ are given in terms of the background metric , with  
\be
P(u)=g^{zz}g_{xx}. \label {pval}
\ee
In effect, eq.(\ref{eqnscal1}) arises from an action
\be
\label{actaa}
S=-\int \sqrt{-g} {1\over 16 \pi G} [ P(u) {1\over 2} g^{uu} (\partial_u Z)^2 - {1\over 2} N(u) g^{tt} (\partial_t Z)^2]
\ee
(we are neglecting the dependence on the spatial $x^i$ coordinates here). 
Using AdS/CFT we can find the response in terms of the canonical momentum
\begin{equation}\label{CNM}
\Pi(u,\omega)=-\frac{1}{16\pi G}\sqrt{-g}P(u)g^{uu}\partial_{u}Z(u).
\end{equation}
The retarded Green's function is then given by the ratio of the response over the source,  
\begin{equation}
 G^{\rm{ret}} = -\frac{\Pi(u,\omega)}{Z(u,\omega)} \Bigg{|}_{u\rightarrow \infty}~~~\label{retardedg}.
\end{equation}
leading to the result from eq.(\ref{kuboa}) 
\begin{equation}
\label{responsefun}
 \eta_{\perp} = \lim_{\omega\rightarrow 0}\frac{\Pi(u,\omega)}{i \omega Z(u,\omega)}\Bigg{|}_{u \rightarrow \infty}.\end{equation}

We now show that the RHS of eq.(\ref{responsefun}) can also be evaluated near the horizon, $u=u_H$, instead of  $u\rightarrow \infty$. 
Since we are interested in the limit  $\omega \rightarrow 0$ we can  
 neglect the second term in eq.(\ref {eqnscal1}) leading to 
 \be
\label{condpi}
\partial_u \Pi=0
\ee
upto $O(\omega)^2$.
This gives 
\be
\label{condphi}
\Pi = C,
\ee
where $C$ is independent of $u$ .
Next, it is easy to see   that there is a solution of   eq.(\ref{eqnscal1}) in the $\omega \rightarrow 0$ limit in which $Z$ is simply a constant. This solution also meets the correct boundary condition at $u \rightarrow \infty$, since,  as can be seen  from eq.(\ref{defperta}),   the   non-normalisable mode must go to a constant at $u \rightarrow \infty$.
Putting all this together we find  that to leading order in the $\omega\rightarrow 0 $ limit both $\Pi$ and $Z$ are constant and thus the ratio in eq.(\ref{responsefun}) being independent of $u$ can also be evaluated at the horizon. 

As a result we get 
\be
\label{membrane}
\eta_{\perp} =\lim_{\omega\rightarrow 0}\,\frac{\Pi(u,\omega)}{i \, \omega \, Z(u,\omega)}\Bigg{|}_{u \rightarrow u_H}.
 \ee

Demanding regularity at the future horizon , we can approximate the behaviour of Z as follows 
\be
\label{clh}
Z \sim e^{-i \omega (t+r_*)},
\ee
where $r_*$ is the tortoise coordinate,
\be
\label{tcoord}
r_*=\int \sqrt{g_{uu}\over g_{tt}} \, du.
\ee

It then follows that
\be
\label{respf}
\eta_{\perp} = \frac{1}{16\pi G}P(u_H)\sqrt{\frac{-g}{g_{tt} g_{uu}}}\Bigg{|}_{u \rightarrow u_H}.
\ee
 The entropy density is
\begin{equation}
\label{ent}
 s=\frac{1}{4 G} \frac{\sqrt{-g}}{\sqrt{g_{uu}g_{tt}}}\Bigg{|}_{u_{H}}.
\end{equation} 

Using the value of P(u) from \eqref{pval} and using eq.(\ref{respf}) and eq.(\ref{ent}) this finally leads to
\begin{equation}
\label{finaletaperp0}
 \frac{\eta_{\bot}}{s}=\frac{1}{4\pi}\frac{g_{xx}}{g_{zz}}\Bigg{|}_{u_{H}}.
\end{equation}

We now see why anisotropic  systems will generically be different from isotropic ones. For an isotropic system rotational invariance makes the ratio ${g_{xx} \over g_{zz}}=1$, leading to the KSS bound, eq.(\ref{kss}). 
However in the anisotropic case in general this ratio will not be unity and thus the  ratio of ${\eta/s}$ can become smaller than ${1\over 4 \pi}$. In the one dilaton system this is what happens leading to the result, eq.(\ref{finaletaperp0}). 
In the rest of this paper we will find many more examples of this type, where anisotropy will allow different metric components to shrink at different rates and attain different values at the horizon, thereby leading to violations of the KSS bound.

\section{Additional examples with anisotropy}
\subsection{Anisotropic solution in two dilaton gravity system}
\label{gravitysolscalar}
To generalise the example in section \ref{rev1dil}, we consider next the case of 
gravity, with a negative cosmological constant,  two massless scalar fields, $\phi_1$ and  $\phi_2$ , both of which we now call dilatons,  in $5$ spacetime dimensions with action,
\be \label{action2scalar}
S_{bulk}=\frac{1}{2\kappa^2}\int d^{5}x\sqrt{-g}~\left(R+12\Lambda-\frac{1}{2}(\partial\phi_{1})^2-\frac{1}{2}(\partial\phi_{2})^2\right).
\ee

%Here $2\kappa^2=16\pi G$ is the gravitational coupling with $G$ being Newton's Constant in 5-dimensions.
Both the dilatons are turned on to be linearly varying, but along different directions:
\be
\label{twodila}
\phi _{1} =\rho_{1}  y ,  \ \ \  \phi _{2} =\rho_{2}  z.  
\ee
The zero temperature near horizon solution is now given by $AdS_3\times R\times R$ (we have set $\Lambda$=1):
\be
 \label{nhz2scalar}
ds^2= -2 u^{2}  dt^{2} + \frac{1}{2 u^{2} } du^{2} + 2 u^{2} dx^{2}+\frac{\rho_{1}  ^{2}}{8  } dy^{2} +\frac{\rho_{2}  ^{2}}{8  } dz^{2}. \\
\ee
% and $ \phi _{1} =\rho_{1}  y ,  \phi _{2} =\rho_{2}  z  $

We see that there are now two different mass scales, $\rho_1, \rho_2$ which characterise the anisotropy. In appendix \ref{Numerics} we show that this near horizon geometry 
interpolates smoothly to asymptotically $AdS_5$. The $SO(2,2)$ symmetry of $AdS_3$ is preserved all along this interpolation. 

At small temperature, $T \ll \rho_1, \rho_2$, the near-horizon solution is given by :
\be
\label{heatednhz2scalar}
ds^2= -2 u^{2}  (1-\frac{\pi^{2}T^{2}}{u^{2}}) dt^{2} + \frac{1}{2 u^{2} (1-\frac{T^{2}\pi^{2}}{u^{2}})  } du^{2} + 2 u^{2} dx^{2}+\frac{\rho_{1}  ^{2}}{8  } dy^{2} +\frac{\rho_{2}  ^{2}}{8  } dz^{2}.
\ee
The horizon lies at 
 \be 
u =u_{h}=  \pi T.
 \ee
The computation of the shear viscosity follows the discussion in \cite{Jain:2014vka} quite closely. The near-horizon $AdS_3$ has $SO(1,1)$ Lorentz invariance in the $t,x$ directions. 
The metric perturbations can be classified in terms of  different spins with respect  to this $SO(1,1)$ symmetry.
The viscosity component $\eta_{xz}$, given by, 
\be
\label{defetperp1}
\eta_{xz} = -\lim_{\omega\rightarrow 0}\frac{1}{\omega}\, Im \big[ G^{R}_{xz,xz}(\omega)\big],
\ee
can be calculated by considering a metric perturbation $Z(u)$ defined so that the full metric with the perturbation takes the form, 
\begin{equation}
\begin{split}
  ds^2=-g_{tt}(u) dt^2+g_{uu}(u)du^2+ &g_{xx}(u)dx^2 + g_{yy}dy^2+g_{zz}(u)dz^2 \\ &+2e^{-i\omega t}Z(u)g_{xx}(u)dx dz.
 \end{split}
\end{equation}
This component has spin $1$ with respect to the $SO(1,1)$ symmetry. It turns out that resulting analysis is quite similar to that in section \ref{viscomp} and this perturbation satisfies an equation of the type given in eq.(\ref{eqnscal1}),
with $P(u)$ given by eq.(\ref{pval}).
The conjugate momentum $\Pi$ is also given by eq.(\ref{CNM}) with $P(u)$ given by eq.(\ref{pval}). As a result $\eta_{xz}$ is given by eq.(\ref{respf}).

The entropy density is given by 
\begin{equation}
 s=\frac{1}{4 G} \frac{\sqrt{-g}}{\sqrt{g_{uu}g_{tt}}}\Bigg{|}_{u_{H}}.
\end{equation} 
This gives, 
\begin{equation}
\label{finaletaperp}
 {\eta_{xz} \over s}=\frac{1}{4\pi}\frac{g_{xx}}{g_{zz}}\Bigg{|}_{u_{H}}.
\end{equation}
which using eq.(\ref {heatednhz2scalar}) becomes
\be
\label{etaperp1}
{\eta_{xz} \over s} =\frac{4\pi T^2}{  \rho_{2}^2}.
\ee
Similarly, for $\eta_{xy}$ we get 
\be
\label{etaperp2}
{\eta_{xy}\over s} = \frac{1}{4\pi}\frac{g_{xx}}{g_{yy}}\Bigg{|}_{u_{H}}=\frac{4 \pi T^2}{  \rho_{1}^2}.
\ee 

We see from eq.(\ref{finaletaperp}), eq.(\ref{etaperp2}) that the relative  ratio of $\eta/s$ for these components is determined by the ratio of  the metric components as one approaches the horizon.

\subsection {Viscosity in the Presence of a Uniform Magnetic Field}
\label{mag}

Here, for completeness, we briefly review a situation where the anisotropy is generated due to a magnetic field which has been studied in considerable depth in \cite{Critelli:2014kra}. We refer to \cite{Critelli:2014kra} for details.
We start with a system with the action 
\begin{equation}
\label{action2gaugefields}
 S = \int d^5 x \sqrt{-g} ( R + 12\Lambda - {1 \over 4} F^{2}   ),
\end{equation}
and consider a solution where the magnetic field 
\be
\label{gf1}
F_{yz}=B,
\ee
with $B$ being  a constant. Such a system was also considered in \cite{ D'Hoker:2009mm}.\\
The resulting near horizon solution  at zero temperature is now again $AdS_3 \times R \times R$, just as in the two dilaton system,  with rotational invariance also preserved in the $yz$ plane. 

The metric is  (we have set $\Lambda$=1)
\begin{align}
\begin{split} \label{qgphor}
ds^2= -  {3} u^{2} dt^{2} + \frac{1}{3 u^{2} } du^{2} +{3} u^{2} dx^{2}+  {1 \over 2\sqrt{3}}|B| dy^{2} +{1 \over 2\sqrt{3}} |B|dz^{2}. \\
\end{split}
\end{align}
The  radius of $AdS_3$,  $R_3^2=1/3 $, in units  where  $\Lambda=1$.

At small temperature, $T \ll B$ the solution is  a black brane in $AdS_3 \times R \times R$ with metric 
\be
\label{heatednhzqgp}
ds^2=-  3u^{2} (1- \frac{c}{u^{2}}) dt^{2} + \frac{1}{ 3u^{2} (1- \frac{c}{u^{2}}) } du^{2} + 3 u^{2} dx^{2}+{1 \over 2\sqrt{3}}  |B| dy^{2} + {1 \over 2\sqrt{3}}|B| dz^{2}, 
\ee

where c is given in terms of T as follows 

\be
c={4 \pi^{2} T^{2}\over 9}.\\
 \ee 
 
The horizon lies at 
 \be 
u =u_{h}= {2\over3} \pi T.
 \ee

The viscosity components $\eta_{xy}=\eta_{xz}\equiv \eta_\perp$. To calculate $\eta_\perp$  we consider
 the  $h_{xz}$ component of  metric perturbation, 
 so that the full metric is of the   form
\begin{equation}
\begin{split}
  ds^2=-g_{tt}(u) dt^2+g_{rr}(u)dr^2+ &g_{xx}(u)dx^2 + g_{yy}(u)dy^2+g_{zz}(u)dz^2 \\ &+2e^{-i\omega t}Z(u)g_{xx}(u)dx dz,
 \end{split}
\end{equation}
%with the gauge field is given by 
%\begin{eqnarray}
%A =  A_{z}  y  dz \\
%\end{eqnarray}
%with the nonzero component of the field strength $ F_{yz}$ is $A_{z}$ .  \\
with $Z(u)$ being  the perturbation that we need to study.  
One can easily show that the other modes decouples from 
$Z(u)$  and so  can be consistently set to  zero.

We find that the resulting analysis is again quite similar to that in section \ref{viscomp} . This perturbation satisfies an equation of the type given in eq.(\ref{eqnscal1}),
with $P(u)$ given by eq.(\ref{pval}).
The conjugate momentum $\Pi$ is also given by eq.(\ref{CNM}) with $P(u)$ given by eq.(\ref{pval}). 

 The resulting value for the viscosity is given by 
 \be
 \label{magv}
 {\eta_\perp\over s}={1\over 4 \pi} {g_{xx}\over g_{zz}}\Bigg{|}_{u_{H}}.
 \ee

%We can heat this solution up to a temperature T given by the following metric

Substituting the  metric components from (\ref {heatednhzqgp}) above we get that 

 \be
\label{etaperpalpha2}
\frac{\eta_{\bot}}{s} = {2\over\sqrt{3}}  \pi \frac{T^{2}}{ |B|}.
\ee

As discussed in \cite{Critelli:2014kra}, this example  may be relevant in the study of QCD, perhaps for heavy ion collisions, and also in the core of neutron stars where strong magnetic fields can arise. 
\subsection{ The Dilaton-Axion System}
\label{oneax}
In the examples considered so far, the near horizon geometry was of the form, $AdS\times R^n$, with the metric components along the $R^n$ directions not contracting as one gets to the horizon. 
It is worth considering other situations where the near horizon geometry is of Lifshitz type instead, with metric components along all the   directions contracting as one approaches the horizon but at different rates. 

An easy way to construct such an example involves a system consisting of gravity with an axion and dilaton with action,
\be \label{action1axionalpha}
S_{bulk}=\frac{1}{2\kappa^2}\int d^{5}x\sqrt{-g}~\left(R+12 \Lambda-\frac{1}{2}(\partial\phi)^2-\frac{1}{2}e ^{2 \alpha  \phi}(\partial\chi)^2\right),
\ee
containing the parameter $\alpha$ which  enters in the dilaton dependence of  the axion kinetic energy term. 
Earlier work in \cite{Rebhan:2011vd}  considered the case with $\alpha=1$. The case $\alpha=-1$ has $SL(2,R)$ invariance. 

 It is easy to see that by turning on a linear profile for the axion one obtains an extremal solution whose near horizon limit is given by  
 ( setting $\Lambda$=1)
\beq
ds^2 = R^{2 } \left(-u^{2}dt^{2} + \frac{du^{2}}{u^{2}}+u^{2}dx^{2} + u^{2}dy^{2}+ \rho^{2}~ u^{\frac{4\alpha ^{2}}{1 + 2 \alpha ^{2}}}dz^{2}\right),\label{met1ax}\\ 
\chi = c_1~\rho ~ z,  \label{ax1} \\
\phi={2\alpha\over 1+ 2 \alpha^2} \log(u),  \label{dil1} \\
 c_1={\sqrt{2 (3+8 \alpha^{2})} \over (1 + 2 \alpha^{2})},\\
 R^{2} = \frac{3+8 \alpha^{2}}{4 + 8 \alpha^{2}}. \label{rad1}
\eeq

This solution breaks rotational invariance along the $z$ direction due to the linearly varying axion, and 
$\rho$ is the mass scale which characterises this breaking of anisotropy. We see that all components of the metric along the spatial directions now shrink as one approaches the far IR, but the rate at which the $g_{zz}$ component vanishes is different from the other spatial components, $g_{xx}, g_{yy}$. 
Let us  also note that for $\alpha=1$ the solution above  agrees with \cite{Azeyanagi:2009pr}.

At small temperature $T \ll  \rho$  the resulting solution has a metric given by
\be
\label{heated1axialpha}
ds^2= R^{2}\left(-u^{2} f(u)dt^{2} + \frac{du^{2} }{u^{2} f(u)}+u^{2}dx^{2} + u^{2}dy^{2}+ \rho^{2}u^{\frac{4\alpha ^{2}}{1 + 2 \alpha ^{2}}}dz^{2}\right),\\
\ee
where $R^2$ is as given in eq(\ref{rad1}) above and 
f(u) is given as 
 \be
 \label{tdef1}
1-\left({16 \pi T \over p^{2} u }\right)^{p},\\
 \ee
 where  p = $\frac{3+8 \alpha^{2}}{1 + 2 \alpha^{2}}$ . 
The axion continues to be linear as in the solution  eq.(\ref{ax1}) and the dilaton is given by eq.(\ref{dil1}). 

The horizon in eq.(\ref{heated1axialpha}) is at 
 \be 
u =u_{h}= { 16 \pi T \over p^{2} }.
 \ee

% \subsubsection{Viscosity in Axion-Dilaton System}
Let us now turn to computing the viscosity.
The shear viscosity component $\eta_{xy}$ satisfies the  KSS bound in eq.(\ref{etaprll1}) . 
Next consider the component $\eta_{xz}=\eta_{yz}$ . To compute this component we can consider the  $h_{xz}$ component of  metric perturbation, 
 so that the full metric is of the   form
\begin{equation}
\begin{split}
  ds^2=-g_{tt}(u) dt^2+g_{uu}(u)du^2+ &g_{xx}(u)dx^2 + g_{yy}(u)dy^2+g_{zz}(u)dz^2 \\ &+2e^{-i\omega t}Z(u)g_{xx}(u)dx dz,
 \end{split}
\end{equation}
where $Z(u)$ is the perturbation that we need to study.  
The dilaton and axion are unchanged and are given by eq (\ref{dil1}) and 
eq (\ref{ax1}) respectively.  
One can easily show that the other modes decouples from 
$Z(u)$  and so  can be consistently set to  zero.

We again find that resulting analysis is similar to that in section \ref{viscomp} and the perturbation satisfies an equation of the type given in eq.(\ref{eqnscal1}),
with $P(u)$ given by eq.(\ref{pval}).
The conjugate momentum $\Pi$ is also given by eq.(\ref{CNM}) with $P(u)$ given by eq.(\ref{pval}). As a result $\eta_{xz}$ is given by eq.(\ref{respf}).

%The above equation in full detail is as follows 
%\be
% Z''(u) + Z'(u) \frac{1}{2}(\frac{g_{tt}'}{g_{tt}}-\frac{g_{uu}'}{g_{uu}} +3 \frac{g_{xx}'}{g_{xx}} + \frac{g_{yy}'}{g_{yy}}- \frac{g_{zz}'}{g_{zz}}) + Z(u) \omega ^{2}\frac{g_{uu}}{g_{tt}}=0
 %\ee

 Thus,  substituting the metric components for the finite temperature solution (\ref {heated1axialpha}) we get 
\be
\label{etaperpalpha1}
\frac{\eta_{\bot}}{s}  = {1\over 4 \pi} {g_{xx}\over g_{zz}}  \sim (\frac{T}{\rho})^{\frac{2}{1+2 \alpha^{2}}}.
\ee 
The dependence on $T$ in eq.(\ref{etaperpalpha1}) follows from the metric eq.(\ref{heated1axialpha}) and the dependence on $\rho$ is then obtained on dimensional grounds. 
Let us note that the temperature $T$ which appears in eq.(\ref{tdef1})
 could be  related to the temperature as measured in the asymptotic $AdS$ coordinates by a rescaling. 
By the asymptotic $AdS$ coordinates we mean those in which the metric takes the standard form:
\be
\label{ads}
ds^2=\bigg[-u^2 dt^2 +{du^2 \over u^2} + u^2 (dx^2 + dy^2 + dz^2)\bigg],
\ee
This is also true for the $x,y$ coordinates in eq.(\ref{heated1axialpha}) and the corresponding coordinates  which appear in eq.(\ref{ads}).  
and also for the $z$ coordinate in eq.(\ref{heated1axialpha}) which  is related to the corresponding coordinate in eq.(\ref{ads})  by a $\rho$ dependent rescaling in general. 
These rescaling factors have to be determined if the coefficient in eq.(\ref{etaperpalpha1}) is to be fixed. 
To do so, one needs to find the full interpolating geometry from the near horizon region, described by eq.(\ref{heated1axialpha}) , to the asymptotic AdS region, eq.(\ref{ads}).

We have carried out such a numerical interpolation for $\alpha=\pm 1$, for which, 
   eq.(\ref{etaperpalpha1}) becomes,
\be
\label{etaperplif1}
\frac{\eta_{\bot}}{s}  \sim ({T \over \rho})^{2/3}. \ee 
We find, within the accuracy of our numerical calculation,  that there is no rescaling of the $T,x,y$ coordinates while the $z$ coordinate is rescaled by a non-trivial $\rho$ dependent factor. 
One consequence is that the temperature $T$ which appears in eq.(\ref{etaperpalpha1}) is the same as the temperature measured in the field theory.

 % 2 axion case with the alpha parameter
 
 \subsection { The two Axion-one Dilaton System }
 \label{twoax}
 For good measure, as another example, we consider a system consisting of gravity with two axions and one dilaton described by the  action 
 \be \label{action2axionalpha}
S_{bulk}=\frac{1}{2\kappa^2}\int d^{5}x\sqrt{-g}~\left(R+12 \Lambda-\frac{1}{2}(\partial\phi)^2-\frac{1}{2}e ^{2 \alpha  \phi}(\partial\chi_{1})^2 - \frac{1}{2}e ^{2 \alpha  \phi}(\partial\chi_{2})^2\right).
\ee

In this case we will see that for a suitable profile for the two axions, the $AdS_4$ symmetry of the near-horizon geometry is broken further to $AdS_3$, with now two of the spatial directions, $y,z$, being characterised by non-trivial Lifshitz exponents. 

The linear profiles for the two axons and resulting near horizon solution is given by (setting $\Lambda$=1)
%It is easy to see that by turning on  linear profiles for the 2 axions one obtains an extremal solution whose near horizon limit is given by
\begin{align}
ds^2 = R^{2 } \left(-u^{2}dt^{2} + \frac{du^{2}}{u^{2}}+u^{2}dx^{2} + \rho^{2}~ u^{\frac{8\alpha ^{2}}{1 + 4 \alpha ^{2}}}dy^{2}+ \rho^{2}~ u^{\frac{8\alpha ^{2}}{1 + 4 \alpha ^{2}}}dz^{2}\right),\label{met2ax}\\ 
\chi  _{1} =  c ~\rho ~ y , \label{axi1}\\
\chi  _{2} =  c ~\rho ~  z , \label{axi2}\\
\phi =\frac{ 4 ~ \alpha \log (u)}{1 + 4\alpha ^{2}}, \label{dil2} \\
c = \frac{ 2}{1 + 4\alpha ^{2}} \sqrt{1+ 8 \alpha ^{2}} , \\
R^{2} = \frac{1+8 \alpha^{2}}{2 + 8 \alpha^{2}} \label{rad2}. 
\end{align}

This metric in this solution has $AdS_3$ invariance, and also a scaling symmetry under which $y,z$ transform with a non-trivial exponent. The linearly varying axions break this scaling symmetry, and also the  rotational invariance along the $y$ and $z$ directions, with   
$\rho$ being  the mass scale which characterise the breaking. 

At small temperature $T \ll  \rho $  the resulting solution has a metric 
\be
\label{heated2axialpha}
R^{2}\left(-u^{2} f(u)dt^{2} + \frac{du^{2} }{u^{2}f(u)}+u^{2}dx^{2} + \rho ^{2}~ u^{\frac{8\alpha ^{2}}{1 + 4 \alpha ^{2}}}dy^{2} + \rho ^{2}~ u^{\frac{8\alpha ^{2}}{1 + 4 \alpha ^{2}}}dz^{2}\right),\\
\ee
where $R^2$ is as given in eq.(\ref{rad2}) above and 
f(u) is given as 
 \be
 \label{tdef2}
1-\left({16 \pi T \over p^{2} u }\right)^{p},\\
 \ee
 where p = $ \frac{2(1+8 \alpha^{2})}{1 + 4 \alpha^{2}} $ .
 
 The two axions continue to be linear as in the solution  eq.(\ref{axi1}), eq.(\ref{axi2}) and the dilaton is given by eq.(\ref{dil2}). 

 The horizon in eq.(\ref{heated2axialpha}) is at 
 \be 
u = u_{h}={ 16 \pi T \over p^{2} }.
 \ee
 
%  \subsubsection{Viscosity in the two Axion-one Dilaton System}
 The $\eta_{xy}$ and $\eta_{xz}$ components of the viscosity are the same., we denote them by $\eta_\perp$.  To calculate these components we consider 
 the  $h_{xz}$ component of  metric perturbation, 
 so that the full metric is of the   form
\begin{equation}
\begin{split}
  ds^2=-g_{tt}(u) dt^2+g_{uu}(u)du^2+ &g_{xx}(u)dx^2 + g_{yy}(u)dy^2+g_{zz}(u)dz^2 \\ &+2e^{-i\omega t}Z(u)g_{xx}(u)dx dz,
 \end{split}
\end{equation}
where  $Z(u)$ is the perturbation that we need to study. \\
The dilaton and axions are unchanged and are given by eq (\ref{dil2}) and 
eq (\ref{axi1}) , eq (\ref{axi2})  respectively. 
One can easily show that the other modes decouples from 
$Z(u)$  and so  can be consistently set to  zero.

As in the previous cases , the analysis here is similar to that in section \ref{viscomp} and this perturbation satisfies an equation of the type given in eq.(\ref{eqnscal1}),
with $P(u)$ given by eq.(\ref{pval}).
The conjugate momentum $\Pi$ is also given by eq.(\ref{CNM}) with $P(u)$ given by eq.(\ref{pval}). As a result $\eta_{xz}$ is given by eq.(\ref{respf}).

Thus,  substituting the metric components for the finite temperature solution (\ref {heated2axialpha}) we get 
\be
\label{etaperpalpha2}
\frac{\eta_{\bot}}{s}  = {1\over 4 \pi} {g_{xx}\over g_{zz}}  \sim \left(\frac{T}{\rho}\right)^{2 \over 1+4 \alpha^{2}}.
\ee 

For the case   $\alpha = \pm 1 $,  eq.(\ref{etaperpalpha2}) becomes,
\be
\label{etaperplif2}
\frac{\eta_{\bot}}{s}  \sim \left({T \over \rho}\right)^{2/5}. \ee 

Interestingly, both in eq.(\ref{etaperpalpha1}) for the one axion case, and in eq.(\ref{etaperpalpha2}) above we see that the maximum value the exponent governing the temperature dependence can take is $2$, and the minimum value, for $\alpha=\infty$, is $0$. 
 
%We also get the same behaviour for the  $h_{xz}$ component of  metric perturbation .

\section {Kaluza Klein Reduction }
\label{kkan}
The previous sections dealt with a number of examples where anisotropic situations gave rise to small values for the viscosity to entropy ratio. 
One common feature of all these examples was that the breaking of isotropy was due to a spatially constant driving force. 
For example, the  dilaton considered  in section \ref{rev1dil}, gives rise to a force proportional to the gradient of the dilaton which is a constant since  the dilaton varies linearly.
One way to see this is by noting that  the stress tensor is no longer conserved and satisfies the equation 
\be
\label{scone}
\partial_\mu <T^{\mu\nu}>~=~<{\hat O}> \partial^\nu \phi,
\ee
as discussed in eq.(6.9) of \cite{Jain:2014vka}. 
Similarly, we consider linearly varying axions in section \ref{oneax} and \ref{twoax}, and a constant magnetic field in section \ref{mag}.

In this section we will present a general argument which should apply to all such situations where the breaking of isotropy occurs due to matter fields which give rise to  a spatially constant driving force. 
We will also assume that  a residual $AdS$ symmetry is preserved in the bulk, and a corresponding Lorentz symmetry is left intact in the boundary theory. 
Fluid mechanics then  corresponds to the dynamics of the goldstone modes associated with the boost symmetries of this Lorentz group. 
The components of the viscosity which give rise to the violation of the KSS bound in the examples considered above   correspond to metric 
perturbations which have spin $1$
with respect to the surviving Lorentz symmetry.  Let  $z$ be a spatial direction in the boundary theory along which there is anisotropy and $x$ be a spatial direction  along which the boost symmetry is left unbroken 
then we will present a general argument below  showing that the viscosity component $\eta_{xz}$, which couples to the $h_{xz}$ component of the metric perturbation. satisfies the relation,
\be\label{mresa}{\eta_{xz}\over s} = {1\over 4\pi} {{ g}_{xx}\over { g}_{zz}}~\Big{|}_{u=u_{h}}. \ee
where ${ g}_{xx}|_{u=u_{h}}, { g}_{zz}|_{u=u_{h}}$ refer to the components of the background metric at the horizon. 
Eq.(\ref{mresa}) is the main result of this section and one of the main results of this paper. We note that it also agrees with all the examples considered above. This result was first obtained for an anisotropic axion-dilaton-gravity system in \cite{Rebhan:2011vd}. An analysis using RG flow and KK reduction, for this system, was carried out in \cite{Mamo:2012sy} along the lines of \cite{Kovtun:2003wp,Iqbal:2008by}. 

For a case with a residual $AdS_{d+1}$ factor in the metric, the  basic idea behind the general analysis will be to consider a dimensionally reduced description, starting from the original 
$D+1$ dimensional theory and going down to the $AdS_{d+1}$ space-time. Different Kaluza Klein (KK) modes in the extra dimensions will not mix with each other since the effects breaking rotational invariance  are in effect spatially constant. 
For example, for cases where there are linearly varying fields, like axions or dilatons, this will be true since the  equations of motion involve only gradients of these fields which are  spatially constant. 
The non-mixing of the KK modes  will greatly ease in the analysis, since we can use the standard formulae of KK reduction and moreover truncate the analysis to the zero modes in the extra dimensions. 
The  off diagonal components of the metric, whose perturbations  carry spin $1$ and which are related to the viscosity components of interest,  will give rise to gauge fields in the dimensionally reduced theory.  By studying the conductivity of 
these gauge fields, which can be related easily to the spin $1$  viscosity components we will derive the result in eq.(\ref{mresa}).

The study of more complicated situations where the breaking of rotational invariance is due to a driving force that also breaks translational invariance is left for the future.

\subsection{The Dimensionally Reduced Theory}

To start, we will consider the case where $D=4$ and $d=3$, so that a residual $AdS_4$ symmetry survives, and the asymptotic geometry, towards the boundary, is $AdS_5$. In this case we start with  $5$ dimensions with a gravitational action :

\be \label{kkh}
S=\frac{1}{2\hat{\kappa}^2}\int d^{5}x\sqrt{-\hat {g}}~(\hat{R} + 12 \Lambda) .
\ee
Here $2\hat{\kappa}^2=16\pi \hat{G}$ is the gravitational coupling with $\hat{G}$ being Newton's Constant in 5-dimension and we set $\Lambda$=1.

Parametrising the $5$ dimensional metric by 
\begin{equation}
\left( \hat{g}_{AB} \right) = \left( \begin{array}{cc}
    e^{-\psi(u)} g_{\mu\nu} +  e^{2 \psi(u)}  A_{\mu} A_{\nu} \; \; & \; \; 
       e^{2 \psi(u)}  A_{\mu} \\
    e^{2 \psi(u)}  A_{\nu} \; \;                                & \; \; 
      e^{2 \psi(u)}
   \end{array} \right) \; \; \; ,
\label{5dMetric}
\end{equation}
and taking all components to be independent of the $z$ direction which we take to be the compactification direction, gives
\begin{eqnarray}
\label{dimredac}
S =\frac{ 1}{2\kappa^2}\int d^{4}x \sqrt{-g} \left( R - {3 \over 2}(\partial\psi)^2  - {e^{3 \psi}\over 4} F^{2} +12 e^{- \psi} \right),
\end{eqnarray}
where we have dropped total derivatives .  \\
We also note that in our choice of parametrisation ,
\be 
\label{gzz}
\hat{g}_{zz}=e^{2\psi}.
\ee
The coefficient of the first term in the matrix in eq.(\ref{5dMetric}) was taken to be $e^{-\psi}$ so that the resulting $4$ dimensional action is in the Einstein frame. 
${\kappa}$ which appears above is related to the $5$ dimensional gravitational coupling ${\hat{\kappa}}$ by 
\be
\label{gn0}
{L \over 2 \hat{\kappa}^2} = {1 \over 2 \kappa^2 },
\ee
where 
$L$   is the length of the compactified $z$ direction. 

So far we have neglected any matter fields. Consider for concreteness the case of the axion-dilaton system considered in section \ref{oneax} with action eq.(\ref{action1axionalpha}) with $\alpha$=1.
Inserting the background solution for the axion 
\be
\label{linax}
\chi=a~ z,
\ee
and taking the dilaton to be independent of $z$ we get from the kinetic energies of the dilaton and axion, 

\begin{eqnarray}
\label{dimredac2}
S =\frac{ 1}{2\kappa^2}\int d^{4}x \sqrt{-g} \left( - \frac{a^{2 }e^{2 \phi }A^{2}}{2 } 
 -\frac{1}{2}(\partial\phi)^2  - {1\over 2}a^{2}e^{2\phi -3 \psi}\right).
\end{eqnarray}
We see that there is an extra term which depends on the gauge field and which gives rise to a mass for it. This term arises due to the linearly varying axion, eq.(\ref{linax}) and is tied to the breaking of translational 
invariance due to this linear variation. 
We see that the terms  in eq.(\ref{dimredac}) and eq.(\ref{dimredac2}) involving the gauge field are quadratic in this field and can be written as 
\be
\label{genac}
S =\frac{1}{2\kappa^2} \int d^4 x \sqrt{-g} \left( {- 1 \over 4 g_{\rm{eff}}^{2}(u)} F^2  -\frac{1}{4} m^{2}(u) A^2 \right),
\ee
where 
\be
m^{2}(u)= 2 a^{2 }e^{2 \phi(u) },
\ee  
and 
\be
\label{effc}
{ 1 \over  g_{\rm{eff}}^{2}(u)} = e^{3 \psi}=\left(\hat{g}_{zz}(u)\right)^{3\over2}.
\ee

The solution in the near horizon region for this dilaton-axion system was given in eq.(\ref{met1ax}) with $\alpha$=1. It is easy to see from this solution that 
\be
\label{depg}
{1\over g_{\rm{eff}}^2}(u) = \rho^3 u^2,
\ee
and 
\be
\label{depm}
m^2(u)= {44\over 9} \rho^{2} u^{\frac{4}{3}},
\ee
and therefore that the gauge coupling and mass vary with the radial coordinate. 

Similarly,  in other cases where there is also a breaking of translational invariance we will  get both a kinetic energy term and a mass term, and in general both the gauge coupling and the mass will vary in the radial direction. 
%Also, more generally, if one reduces on more than one direction, we will get additional gauge fields. with  varying gauge couplings and masses along the radial direction. 
For the subsequent analysis we  will analyse the perturbations of the gauge fields in the $4$ dimensional theory given in eq.(\ref{genac}).
%resulting lower dimensional theory which we take to be in $d+1$ dimensions with an action
%\be
%\label{gendac}
%S =\frac{1}{2\kappa^2} \int d^{d+1} x \sqrt{-g} \left( {- 1 \over 4 g_{\rm{eff}}^{2}(u)} F^2  -\frac{1}{4} m^{2}(u) A^2 \right)
%\ee
%with a varying gauge coupling in general, and a mass term $m^2$. 
Such a system was considered in \cite{Myers:2009ij,Chakrabarti:2010xy} and our subsequent discussion  closely follows this reference. 
As we will see later, the conductivity of 
these gauge fields can be related easily to the spin $1$  viscosity components using which we will derive the result in eq.(\ref{mresa}).
 Let us mention for now that the  essential reason for this is that the two-point correlator  of the current operator gives  the conductivity of the gauge field, while the two-point stress tensor in the higher dimensional theory is related to the viscosity.
 Since the gauge field is obtained from the spin $1$ component of the metric in the higher dimensional theory, these two correlators are closely related.

The $3+1$ dimensions, include time, $t$, the radial direction $u$, and additional space directions, one of which we denote by $x$. 
To study the conductivity we consider a  perturbation for the $x$ component of the gauge field,
\begin{equation}
\label{gans}
{ A}_{x} (\vec x,t,u) = \int{ d\omega d^{3}\vec k \over (2 \pi)^{4} } e^{-i \omega t + \vec k.\vec x}Z(u,\omega) .\quad
\end{equation}
This gauge field  perturbation decouples from the rest (we have set perturbations of the axion to vanish even before the KK reduction in the example above, this turns out to be  a consistent thing to do). 
 $Z(u,\omega)$ satisfies the equation
\begin{equation}
\frac{d}{du}(N(u) \frac{d}{du}Z(u,\omega))-\omega^2 N(u)~ g_{uu} g^{tt} Z(u,\omega)+  M(u) Z(u,\omega)=0,
\label{eqnmotion}
\end{equation}
 
with 
\be
\label{defN}
N(u) = \sqrt{-g}\frac{1}{g_{\rm{eff}}^2}g^{xx}g^{uu},
\ee
 and 
 \be
 \label{defM}
 M(u) =- {m^{2}(u) \sqrt{-g} \over 2~ g_{xx}}. 
 \ee

%The action eq.(\ref{gendac}) in terms of $Z$ and $Z^{*}$ becomes
%\be
%\label{genz}
%S =\frac{1}{2\kappa^2} \int d^{d+1} x  \left(-{N(u)\over 2}(\partial_{u}A_{1}+\partial_{u}A_{2})^{2} +\frac{1}{2} M(u){\cal A}_{x}^2 -{\omega^{2}g^{tt}g_{uu}\over 2}N(u)(A_{1}-A_{2})^{2}\right)
%\ee
Treating the radial coordinate $u$ as the analogue of time we can read off the 
``momentum'' conjugate to $Z$ from eq.(\ref{genac}) to be 
 \be
 \label{pidef}
 \Pi(u,\omega) = {\delta S \over \delta Z'(u,-\omega)}
=-\frac{1}{2 \kappa^{2}} N(u) Z'(u,\omega) ,
\ee where $Z' = \frac{d}{du} Z(u,\omega) $ and N(u) as given in eq(\ref{defN}).

The conductivity is given by 
\be
\label{defsig}
\sigma(u,\omega) = {\Pi (u, \omega)  \over i \omega Z(u, \omega) } \big{|}_ {u\rightarrow \infty, \omega \rightarrow 0},
\ee
where $Z$ and $\Pi$ are  the asymptotic values of the perturbation and conjugate momentum defined in eq.(\ref{pidef}) in the region $u \rightarrow \infty$.

We assume that the underlying higher dimensional geometry is  asymptotically $AdS_5$ space and that the back reaction due to the matter fields which break the rotational invariance
dies out compared to the cosmological constant in this asymptotic region. This is true in all the examples studied above where the geometry becomes $AdS_5$ when $u \rightarrow \infty$. 
It is then easy to check, as discussed in appendix \ref{bdb} that the ratio on the RHS in eq.(\ref{defsig}) becomes independent of $u$ when $u\rightarrow \infty$. 

We can write $\sigma(u,\omega)$ as the sum of real and imaginary parts as $\Re(\sigma(u,\omega))+ i~ \Im(\sigma(u,\omega))$. We will be interested in the real part $\Re(\sigma)$ since that is related to the viscosity components of interest. It is easy to see from our definition, eq.(\ref{defsig}) that 
\begin{eqnarray}
 \Re(\sigma(u,\omega))  &=& \Im\Bigg(\frac{\Pi(u,\omega)Z(u,-\omega)}{ \omega Z(u,\omega) Z(u,-\omega)}\Bigg)\big{|}_ {u\rightarrow \infty,~ \omega \rightarrow 0}.
\end{eqnarray}
where $\Pi(u,\omega)$ is defined in eq.(\ref{pidef}).\\
To evaluate the   RHS in the limit $\omega \rightarrow 0$, it will be sufficient to consider the leading  order behaviour of the denominator. 
Since $Z(u,\omega)$ is real to leading order when $\omega \rightarrow 0$ we obtain 
%The leading behaviour for $Z(u, \omega)$ as $\omega \rightarrow 0$  will suffice evaluate the RHS. This leading order behaviour for $Z(u)$ is real. 
%Thus eq.xx can be written as 
\be
\label{exp22}
\Re(\sigma)= \frac{\Im(\Pi(u,\omega)Z(u,-\omega))}{ \omega ~Z^2(u)}\big{|}_{u\rightarrow \infty,~ \omega\rightarrow 0}.
\ee

%As shown in appendix \ref{b2hz}
 The numerator of RHS of eq.(\ref{exp22})  is independent of $u$ (appendix \ref{b2hz}) and can therefore be evaluated at $u=u_h$ instead of $u \rightarrow \infty$.  After some more  simplification this gives
 \be
\label{exp33}
\Re(\sigma) =  \sigma_{H}~~\Bigg(\frac{Z(u_{h})}{Z(u\rightarrow \infty)}\Bigg)^{2}\big{|}_{ \omega\rightarrow 0} ,
\ee 
where $\sigma_{H} $ is the conductivity evaluated at the horizon and its expression is given by,
\begin{equation}
\label{conh}
\sigma_{H} = \frac {1}{2 \kappa^{2}g_{\rm{eff}}^2} ~ \Big{|}_{u=u_{h}}.
\end{equation}
See appendix \ref{b2hz} for more details. 

To proceed we need to evaluate the ratio $Z(u_h)\over Z(u\rightarrow \infty)$. 
For this purpose we go back to the underlying higher dimensional theory with which we started in which the gauge field is actually an off diagonal component of the metric, eq.(\ref{5dMetric}). 
The background about which we are calculating the behaviour of the perturbation is diagonal in the metric with all components being only a function of $u$. Now consider  a coordinate transformation
 $ x\rightarrow x + \alpha z $, with all the other coordinates remaining the same. It is easy to see that under this transformation the metric now acquires an off-diagonal component
 \be
 \label{odm}
 \delta \hat{g}_{xz}= \alpha {\hat g}_{xx},
 \ee
 with all the other components of the background metric staying the same.  Note that in our notation the hatted metric refers to the $5$ dimensional one while the unhatted metric refers to the $4$ dimensional Einstein frame metric, see eq.(\ref{5dMetric}). 
 
 Since we have merely carried out a coordinate transformation it is clear that $\delta {\hat g}_{xz}$ in eq.(\ref{odm}) must satisfy the equations of motion for small perturbations about the starting background. 
 Comparing with eq.(\ref{5dMetric}) we find that this corresponds to turning on a gauge field
 \be
 \label{defaxp}
 A_x = \alpha {\hat{g}_{xx}\over e^{2\psi}},
 \ee
 which must therefore solve the equation (\ref{eqnmotion}) in the limit $\omega \rightarrow 0$ with 
 \be
 \label{valza}
 Z(u)= \alpha {\hat{g}_{xx}\over e^{2\psi}}. 
 \ee
 %Note that here $g_{xx}, e^{2\psi}$ correspond to the values of the $g_{xx}, g_{zz}$ components of the metric in the
 % background geometry. 
  In this way we can exploit the co-ordinate invariance of the underlying higher dimensional theory to obtain a solution for $Z(u)$ in the $\omega \rightarrow 0$ limit. More over it is easy to see that this solution meets the correct boundary condition at $u \rightarrow \infty$. As was mentioned above, we are assuming that the higher dimensional metric is asymptotically $AdS_5 $ space. The ratio ${\hat{g}_{xx} \over e^{2 \psi}}$ therefore goes to unity and $Z(u)$ goes to a constant  which is the correct behaviour needed, as is also discussed
  in appendix \ref{bdb}. 
  
  With the solution  eq.(\ref{valza}) at hand we can now evaluate the ratio  $Z(u_h)\over Z(u \rightarrow \infty)$. The arbitrary constant $\alpha$ drops out and we get that 
\be
\label{valrat}
{Z(u_h) \over Z(u\rightarrow \infty)}= {\hat{g}_{xx}\over e^{2\psi}}\Big{|}_{u=u_{h}}.
\ee
  
  Substituting  in eq.(\ref{exp33}) and using eq.(\ref{conh}) we  get that the conductivity is given in terms of $\rm g_{eff}^2$ and various metric opponents at the horizon by 
  \begin {equation}
\label{conductivity2}
\sigma=\frac {1}{2 \kappa^{2}g_{\rm{eff}}^2}~ \left(\frac{\hat{g}_{xx}}{e^{2\psi}}\right)^{2}\Big{|}_{u=u_{h}}.
\end {equation}
From eq.(\ref{effc})  and using eq.(\ref{gzz})  from our parametrisation eq.(\ref{5dMetric}) , we finally get that 
\be
\label{fsigma}
\sigma = \frac {1}{2 \kappa^{2}} {\hat{g}_{xx}^{2} \over \sqrt{\hat{g}_{zz}}}.
\ee

   Note that we have been able to obtain an expression independent of  $m^2$ that only depends on the metric components $\hat{g}_{xx}, \hat{g}_{zz}$ in the $5$ dimensional theory. In the subsequent discussion we somewhat loosely denote  $Re(\sigma)$ by $\sigma$ itself.
   
   \subsection{The Viscosity To Entropy Ratio}
   \label{vent}
   
   The next step is to relate the conductivity obtained above to the viscosity. 
   This is in fact straightforward. Kubo's formula relates the components of the viscosity to the two point function of corresponding  components of the stress tensor $T_{ij}$  in eq.(\ref{kuboa}). 
   This two point function is obtained by calculating the response to turning on suitable metric perturbations in the bulk. 
   We will be assuming, as was mentioned above, that asymptotically the background metric is $AdS_5$. Thus as $u \rightarrow \infty$, $\hat{g}_{\mu\nu} \rightarrow u^2 \delta_{\mu\nu}$
   for all components other than along the $u$ direction, as discussed in appendix \ref{bdb}. 
   %Denote the components of the background metric which is asymptotical $AdS_5$ by $g_{(0)\mu\nu}$. Asymptotically when $u \rightarrow \infty$, $g_{(0) \mu\nu}\sim u^2$.
   The off - diagonal metric perturbations required for the shear viscosity  then behave like 
   $$\delta \hat{g}_{\mu\nu}= u^2 h_{\mu\nu}$$
   as $u \rightarrow \infty$, where $h_{\mu\nu}$ is independent of $u$. 
   %On comparing with eq.xx we see that $h_{zx}= A_x$.
   The viscosity  component $\eta_{xz}$ is then given by 
   \be\label{etaaa}\eta_{xz}=-{1\over \omega} \Im \left(<T_{xz}(\vec{ k_1}, \omega) T_{xz} (\vec{k_2}, \omega ) >'\right)\big{|}_{\vec{k_1}, \vec{k_2} \rightarrow 0, \omega \rightarrow 0},\ee
   where the prime subscript on the RHS means that the overall energy momentum conserving delta function has been removed. 
   From AdS/CFT we have that 
   \be 
   <T_{xz}(\vec{k_1}) T_{xz}(\vec{k_2})> ~= {\delta^{2}  S \over \delta h_{xz}(\vec{k_1})
    \delta h_{xz}(\vec{k_2})}.
   \ee

   The conductivity in an analogous way is given by 
   \be\label{sigmaaa} \sigma= -{1\over \omega} \Im \left(<J_x (\vec{k_1}, \omega ) J_x(\vec{k_2}, \omega )>'\right)\big{|}_{\vec{k_1}, \vec{k_2} \rightarrow 0, \omega \rightarrow 0},\ee
   which in turn  can be calculated from the bulk response since 
   \be 
   <J_x (\vec{k_1} ) J_x(\vec{k_2})>~= {\delta^{2} S \over\delta A_{x}(\vec{k_1}) ~\delta A_{x}(\vec{k_2})}.
   \ee

   On comparing with eq.(\ref{5dMetric}) we see that the zero mode of $h_{zx}$ in the $z$ direction is in fact $A_x$.  This shows  that   $\eta_{xz}$ and $\sigma$ are essentially the same upto one minor factor of $L$ the size of the $z$ direction. 
   This factor arises because the prime subscript in eq. (\ref{etaaa}) and eq.(\ref{sigmaaa}) are different, in the first case  the momentum conservation delta function removed includes a delta function in the $z$ direction, whereas in the case of the conductivity it does not include this delta function. Accounting for the difference gives
   \be
   \label{viscon}
   \eta_{xz}= {\sigma \over L}.
   \ee
   The entropy density in the $5$ dimensional theory is given by 
 \be
 \label{entd0} 
 s={2\pi\over \hat{\kappa}^{2}} A= {2\pi\over \hat{\kappa}^{2}} \sqrt{\hat{g}_{xx} \hat{g}_{yy}\hat{g}_{zz}},
 \ee
(this is also the same as the entropy density in the $4$ dimensional theory divided by $L$). 
From eq.(\ref{viscon}) , eq.({\ref{conductivity2}), eq.(\ref{gzz}), eq.(\ref{entd0})and eq.(\ref{gn0}), we can now write the ratio
\be 
\label{fvis}
{\eta_{xz}\over s }= {{\sigma \over L}\over s }= {1 \over 4\pi}{\frac {1}{g_{\rm{eff}}^2}~ \left(\frac{\hat{g}_{xx}}{\hat{g}_{zz}}\right)^{2}\over\sqrt{\hat{g}_{xx} \hat{g}_{yy}\hat{g}_{zz}}   }\Big{|}_{u=u_{h}}.
\ee
Using  eq.(\ref{effc}), eq.(\ref{gzz}) in the above expression and using isotropy along $x$ and $y$, we arrive at the following result
\be 
\label{vis2}
{\eta_{xz}\over s }= {1 \over 4 \pi} {\hat{g}_{xx}\over \hat{g}_{zz}}\big{|}_{u=u_{h}}.
\ee
This general result agrees with the ones we obtained in all the examples we studied in the previous sections. We see that independent of the details of the matter fields which were responsible for the breaking of the rotational symmetry we get a general result in eq.(\ref{fvis}).
This result shows that when the ratio of the metric components ${\hat{g}_{xx}\over \hat{g}_{zz}}$ at the horizon becomes smaller than unity the KSS bound will be violated. 

\subsection{Generalisation To Case with Additional Directions}
In the preceding discussion of this section we have considered the dimensional reduction from $5$ to $4$ dimensions. However, it is easy to generalise these results for the case where we start with $D+1$ dimensions and KK reduce to $d+1$ dimensions. In fact, this generalisation is needed for the situation discussed earlier with a magnetic field where the residual symmetry arises due to an $AdS_3$ factor instead of an $AdS_4$ in the geometry. Our analysis closely follows \cite{Maharana:1992my}. The dimensional reduction in this case  will give rise to $D-d$ gauge fields . \\
Following \cite{Maharana:1992my}, we parametrize the higher dimensional metric as :
\begin{equation}
\left( \hat{g}_{AB} \right) = \left( \begin{array}{cc}
     g_{\mu\nu} +   A_{\mu}^{(1)\gamma}  A_{\nu \gamma}^{(1)} \; \; & \; \; 
         A_{\mu \beta}^{(1)} \\
         A_{\nu \alpha}^{(1)} \; \;                                & \; \; 
     G_{\alpha \beta}
   \end{array} \right) \; \; \; ,
\label{hdmet}
\end{equation}

where the $D+1$ dimensional vielbein is given by
\begin{equation}
\left( \hat{e}^{\hat{r}}_{\hat{\mu}} \right) = \left( \begin{array}{cc}
     e^{r}_{\mu} \; \; & \; \; 
         A_{\mu }^{(1)\beta}E_{\beta}^{a }\\
         0 \; \;                                & \; \; 
         E_{\alpha}^{a}
   \end{array} \right) \; \; \; ,
\label{vb}
\end{equation}
with $G_{\alpha \beta} = E^a_{\alpha} \delta_{ab} E^b_{\beta}$ and $g_{\mu \nu} = e^r_{\mu}\eta_{rs} e^s_{\nu}$. Here $\alpha,\beta=1,..D-d$ denote the directions over which the reduction has been carried out and $\mu,\nu=0, 1, \cdots d$ are the ones left in the lower dimensional theory. It also follows from the parametrisation that 
\be 
\sqrt{- \hat g} = \sqrt{-g} \sqrt{ G},  
\ee
where G is the determinant of the internal metric $ G_{\alpha \beta}$. Additional matter fields required for breaking rotational invariance which also break the translational invariance in the compactified directions give mass terms for the gauge fields, which will vary in general  in the radial direction. Neglecting these additional matter fields for now we start with the action
%However, as seen in the last section , the matter fields in the action that break the rotational symmetry do not affect the general results involving conductivity and viscosity . In fact, the lesson we learnt there is to carefully keep track of the effective gauge coupling of the field strength that arises from the gravitational sector of the higher dimensional action. This motivates us to start with the action in $D+1$ dimensions as given below-
$$S_{\hat{ g}} ={1\over 2 \hat{\kappa}^{2}} \int d^{D+1}x~ \sqrt{- \hat {g}}~
\big [\hat{ R} +\Lambda\big ]$$
As shown in \cite{Maharana:1992my} the dimensionally reduced action in $d+1$ dimensions becomes
\be
\label{acdd}
\begin{aligned}
 S = \frac {1}{2 \kappa^{2}}\int &d^{d+1}x \sqrt {-g}~ e^{-\phi}\left( R +\Lambda + g^{\mu \nu}
\partial_{\mu} \phi \partial_{\nu} \phi+{1 \over 4} g^{\mu \nu} \partial_{\mu} G_{\alpha \beta} \partial_{\nu}
G^{\alpha \beta}\right.\\
&\left. - {1 \over 4} g^{\mu \rho} g^{\nu \lambda} G_{\alpha
\beta} F^{(1)\alpha}_{\mu \nu} F^{(1) \beta}_{\rho \lambda} \right),
\end{aligned}
\ee
where 
\be 
\label{indet}
\phi =  - {1 \over 2} {\rm log~ det}\, (G_{\alpha \beta}) \Rightarrow e^{-\phi}=\sqrt{G},
\ee
 where G is the determinant of the internal metric $G_{\alpha\beta}$,

\be
F_{\mu \nu}^{(1) \alpha} = \partial_{\mu} A_{\nu}^{(1) \alpha} -
\partial_{\nu} A_{\mu}^{(1) \alpha}, 
\ee
and 
${\kappa}$ which appears above is related to the $5$ dimensional gravitational coupling ${\hat{\kappa}}$ by 
\be
\label{gn}
{L^{D-d} \over 2 \hat{\kappa}^2} = {1 \over 2 \kappa^2 },
\ee
where 
$L^{D-d}$   is the volume  of the compactified directions .

For simplicity we assume that the internal metric $G_{\alpha \beta} $ is diagonal and focus on the $\hat{g}_{xz}$ component of the metric perturbation (where x represents a  spatial direction along which the boost symmetry is left unbroken and z represents an anisotropy direction in the boundary field theory). Comparing the last term in the action eq.(\ref{acdd}) with the kinetic energy term,  $\sqrt{- g} \left( {- 1 \over 4  g_{\rm{eff}}^{2}(u)} F^2\right)$, as given in eq.(\ref{genac}), we then find the effective gauge coupling, for the corresponding gauge field $A$ is 
\be 
\label{effcd}
{ 1 \over   g_{\rm{eff}}^{2}}= e^{-\phi} g_{zz}. 
\ee

As mentioned above,  additional matter fields give rise to mass terms for the gauge fields. We will also take these mass terms to be diagonal for simplicity.
The resulting equation for the $x$ component of the gauge field $A_x$ is then of the form given in eq.(\ref{eqnmotion}), where we have expanded $A_x$ as given in eq.(\ref{gans}). 
It can then be argued (see Appendix \ref{b2hz} for details)  that the conductivity in the lower $d+1$ dimensional theory \footnote{ With our choice, eq.(\ref{hdmet}), the  dimensional reduction  results in an action  which is not in Einstein frame. We could have performed a conformal transformation to bring the lower dimensional action back to the Einstein frame. Our end result however will be independent of this choice.} is given by 
\begin{eqnarray}
\label {conexp}
\Re(\sigma) &=& \frac {1}{2\kappa^{2}} \Bigg(\sqrt{\frac{\ g_{uu}}{\ g_{tt}}}  \ N(u)\Bigg)_{u=u_{h}} \Bigg(\frac{ Z(u_{h})}{ Z(u\rightarrow \infty)}\Bigg)^{2} \nonumber\\
 &=& \frac {1}{2 \kappa^{2}}\Bigg(\sqrt{\frac{\ g_{uu}}{\ g_{tt}}}  ~\sqrt{-\ g}\frac{1}{\ g_{\rm{eff}}^2}\ g^{xx}\ g^{uu} \Bigg)_{u=u_{h}}\Bigg(\frac{ Z(u_{h})}{ Z(u\rightarrow \infty)}\Bigg)^{2}\nonumber
 \end{eqnarray}
 Thus we find
 \be 
 \label{condgen}
 \Re(\sigma)= \sigma_{H}~~\Bigg(\frac{ Z(u_{h})}{ Z(u\rightarrow \infty)}\Bigg)^{2}, 
\ee
where $\sigma_{H} $ is the conductivity evaluated at the horizon and its expression is given by,
\begin{equation}
\sigma_{H} = \frac {1}{2 \kappa^{2}\ g_{\rm{eff}}^2}~ { g_{xx}^{d-1 \over 2} \over g_{xx}}\Big{|}_{u=u_{h}},\end{equation}
where we have used isotropy along the spatial directions (besides $u$) in the lower dimensional theory.
Using eq.(\ref{condgen}), eq.(\ref{effcd}), eq.(\ref{indet}) we get
\begin{eqnarray}
\Re(\sigma)= \frac {1}{2 \kappa^{2} g_{\rm{eff}}^2}~ { g_{xx}^{d-1 \over 2} \over \ g_{xx}}\Big{|}_{u=u_{h}}~~\Bigg(\frac{ Z(u_{h})}{ Z(u\rightarrow \infty)}\Bigg)^{2} \nonumber \\
 =\frac {1}{2 \kappa^{2}}~e^{-\phi} g_{zz}  { g_{xx}^{d-1 \over 2} \over  g_{xx}}\Big{|}_{u=u_{h}}~~\Bigg(\frac{Z(u_{h})}{Z(u\rightarrow \infty)}\Bigg)^{2}\nonumber\\
 =\frac {1}{2 \kappa^{2}}\sqrt{G}~ g_{xx}^{d-1\over2} {g_{zz} \over g_{xx}}\Bigg(\frac{Z(u_{h})}{Z(u\rightarrow \infty)}\Bigg)^{2}.\nonumber\\
\end{eqnarray}
We can now repeat the analysis done in the previous section to evaluate the ratio $Z(u_h)\over Z(u\rightarrow \infty)$, by using general coordinate invariance in the  underlying higher dimensional theory and noting that  the gauge field is  an off-diagonal component of the metric, eq.(\ref{hdmet}) (for details see eq.(\ref{valrat})).\\
Thus we get
\begin{eqnarray}
\Re(\sigma)=  \frac {1}{2 \kappa^{2}}\sqrt{G}~ g_{xx}^{d-1 \over 2} {g_{zz} \over g_{xx}}Z(u_{h})^{2}\Big{|}_{u=u_{h}}\nonumber \\
= \frac {1}{2 \kappa^{2}}\sqrt{G} ~g_{xx}^{d-1\over 2} {g_{zz} \over g_{xx}}~{g_{xx}^{2}\over g_{zz}^{2}}\Big{|}_{u=u_{h}}\nonumber\\
= \frac {1}{2 \kappa^{2}}\sqrt{G} ~g_{xx}^{d-1 \over 2} ~{g_{xx}\over g_{zz}}\Big{|}_{u=u_{h}}.
\end{eqnarray}

The higher dimensional entropy density is 
\be 
\label{entd}
s={2 \pi \over \hat{\kappa}^{2} } \sqrt{G} ~ g_{xx}^{d-1 \over 2}.
\ee
Hence we  arrive at the result 
\be 
{\sigma \over s}= L^{D-d} {1\over 4 \pi }{g_{xx}\over g_{zz}}\Big{|}_{u=u_{h}}.
\ee

Finally, the arguments given in subsection \ref{vent} allows us to connect $\eta_{xz}$ computed in the higher dimension to $\sigma$ in the following way 
\be
\eta_{xz}= {\sigma \over L^{D-d}}.
\ee
Thus we find 
\be
{\eta_{xz}\over s} = {{\sigma \over L^{D-d}}\over s} ={1\over 4 \pi }{g_{xx}\over g_{zz}}\Big{|}_{u=u_{h}},
\ee
which agrees with  the examples we have studied in the previous sections.

\section{Conclusion}
\label{cncl}
In this paper we have considered a variety of anisotropic examples, and have shown that suitable components of the viscosity can become very small in the highly anisotropic case and can parametrically violate the bound, eq.(\ref{kss}).
All our examples have the feature that the breaking of rotational invariance is due to an externally imposed forcing function which is translationally invariant. E.g.  due to linearly  varying  scalars which give rise to a constant forcing function, or due to a spatially constant magnetic field, which was studied earlier in \cite{Critelli:2014kra}. Another common feature in all our examples is that some residual  Lorentz  symmetry  survives at zero temperature. In the second half of the paper  we show in considerable generality that for all cases with these two features, the   components of the viscosity tensor, which correspond to metric perturbations which carry spin $1$ with respect to the unbroken Lorentz symmetry, satisfy the relation eq.(\ref{mresaa}). In the anisotropic case the ratio of the metric components on the RHS of eq.(\ref{mresaa})  can become very small as $T \rightarrow 0$, resulting in a parametrically large violation of the KSS bound. This is indeed true for  the examples we consider, all of which satisfy eq.(\ref{mresaa}) .

Besides allowing for a computation of the viscosity with relative ease, the gravitational description also  provides an intuitive  understanding of  why such violation of the KSS bound may arise.  In the absence of isotropy the different metric perturbations break up into components with different values of spin with respect to the remaining Lorentz symmetry. Spin $2$ components, if present, give rise to viscosity coefficients which satisfy the KSS bound.
But  spin $1$ components can violate it. In fact the spin $1$ components are akin to gauge fields, and the corresponding calculations for these  components of the  viscosity therefore becomes similar to those for  conductivity. These are well known   in several AdS/CFT examples, and also in nature,  to sometimes become very  small.

In weakly coupled theories, with well defined quasi particles, we would expect, \cite{Landau1987Fluid}, \cite{REI65}, that
\be
\label{bwc}
{\eta \over s} \sim {l_{mfp}\over \lambda_{dB}},
\ee
where $l_{mfp}, \lambda_{dB}$ refer to the mean free path and the de Broglie wave length for the quasi particles. 
This leads to the intuitive expectation that at strong coupling the ratio $\eta/s\sim O(1)$. However, here we see that at strong coupling, where the gravity description is valid, some components of the viscosity tensor in the anisotropic case violate this relation and can become parametrically smaller.

The generality of our result suggests the  possibility that this behaviour might  happen in nature too. 
It would be very exciting if this  can be probed in experiments,  perhaps on cold atom systems, or in  QCD. 

Ordinarily, QCD at finite temperature is described by a  homogeneous and isotropic phase for which  the calculations discussed here are not relevant.
This is true even when we consider situations which come about  due  to anisotropic initial conditions, as might arise in heavy ion collisions. The behaviour of the QCD fluid   in these situations is still  governed by  rotationally invariant Navier Stokes equations with appropriate viscosity coefficients.  However, this could change if a sufficiently big magnetic field   is turned on breaking rotational invariance \footnote{A magnetic field of order $10^{16}$ Tesla or so is needed in order to contribute an energy density comparable to the QCD scale $\sim 200$ Mev.}.  The resulting  equilibrium phase  could then be highly anisotropic and our results, and earlier work, \cite{Critelli:2014kra}, hint that suitable components of the viscosity might   become small. It has been suggested that such an intense magnetic field might perhaps arise in the interior of some highly magnetised neutron stars \footnote{We thank  Gergely Endr{\"o}di and Gunnar Bali for a discussion on this issue.},  see \cite{Bocquet:1995je}, \cite{Harding:2006qn} and \cite{Endrodi:2014lja}. It has  also been suggested that  strong magnetic fields might  actually  arise in the highly relativistic heavy ion collisions (see  \cite{Kharzeev:2007jp}, \cite{Fukushima:2008xe} and \cite{Skokov:2009qp}), although in this case the transitory nature of these fields must also then be taken into account.    

Turning to cold atom systems, the unitary Fermi gas has also been observed to have a value of $\eta/s$ close to the KSS bound. Perhaps some way to introduce the breaking of rotational invariance can be found in this system.
 It would then be very interesting to examine the resulting behaviour of the viscosity tensor. Even at small anisotropy one might be able to see a trend where some  components start getting smaller than the bound. A natural way to incorporate anisotropy in this case might be to consider the effects of an asymmetric trap \footnote{We thank Mohit Randeria for  very helpful discussions in this regard and also for his comments about the spin diffusion experiments.} . 
 
 It is worth mentioning that   the spin $1$ viscosity components, which become very small in our work,  govern the diffusion of the  momentum  components oriented transverse to the direction in which the initial inhomogeneity is set up. For example, take a case with anisotropy in the $z$ direction. If  the momentum along the $x$ direction, $p_x$, is now taken to have an initial gradient along the $z$ direction, then its diffusion is governed by the viscosity component $\eta_{xz}$, with diffusion length
\be
\label{dl}
D_{\perp}={\eta_{xz}\over s T},
\ee
where $s$ is the entropy density. A  small value of ${\eta_{xz}\over s}$ then gives rise to a small value for the diffusion constant \footnote{The anisotropy force  in this case would  act in the $z$ direction.  This force does not directly enter in the diffusion equation for 
$p_x$. For significant anisotropy, $\rho/T\gg  1$,   the force is big,  and as a result the fluid cannot move in the $z$ direction at all. This follows from the bulk geometry, e.g. $AdS_4\times R$ in the case considered in section \ref{rev1dil}, where Lorentz invariance along the $z$ direction is manifestly broken.}$D_{\perp}$ in units of temperature.

It is perhaps worth mentioning in this context  that there have been some recent measurements of spin diffusion in the unitary fermi gas system \footnote{We thank Sean Hartnoll for bringing these experiments to our notice.}. In three space dimensions, with rotational invariance intact, the transverse spin diffusion constant is measured to be close to the bound which arises from standard Boltzmann transport theory based on quasi particles, see \cite{2014Sci...344..722B}. However, in a quasi-two space dimensions \cite{2013NatPh...9..405K}, it was found that the transverse spin diffusion constant is about three orders of magnitude smaller than this bound. It would be worth exploring if these observations can be  related to the results presented here. 

We have not analysed the stability of the anisotropic solutions discussed in this paper in any detail. For the one dilaton case this question was analysed at considerable length in \cite{Jain:2014vka} and no instabilities were found. 
This suggests that some  examples studied here, e.g., the two dilation case, also could be stable. We leave a more detailed analysis of this question for the future. It is worth noticing that if an instability appears, it will be 
when the temperature $T \sim \rho$, where $\rho$ is the scale of the anisotropy. As a result one expects $\mathcal{O}(1)$ violations of the bound for such systems as well, although not  violations where the viscosity becomes parametrically small. 
On a more theoretical note, it would be worth obtaining  string theory embeddings of the anisotropic systems we have studied here and examining if they are  stable. Some embeddings for the axion dilaton system were studied in \cite{Azeyanagi:2009pr} and for the one dilaton case in \cite{Jain:2014vka} and were found to be unstable, since they contained fields which lay below the BF bound of the near
horizon geometry. In another instance, e.g. \cite{Polchinski:2012nh}, though, a  stable supersymmetric system with anisotropy was found where suitable components of the viscosity become vanishingly small at low temperatures, just as in our analysis here.

We have discussed situations where the breaking of rotational invariance is explicit, due to an externally applied source. It would also be interesting to extend this analysis to cases where the breaking is spontaneous.
Examples are known on the gravity side of such phases in the literature, see, e.g., \cite{Domokos:2007kt,Rozali:2007rx,Nakamura:2009tf,Ooguri:2010kt,Ooguri:2010xs,Bergman:2011rf,Donos:2011qt,Donos:2011ff,Donos:2012gg,Donos:2012wi}.
Another direction is to consider  Bianchi spaces which have been discussed in \cite{2012JHEP...07..193I,Iizuka:2012pn}, and which describe homogeneous but anisotropic phases in general. 
Some discussion of transport coefficients in such phases using the gravity description can be found in \cite{Ovdat:2014ipa}.

\section{Acknowledgements}
We thank Gunnar Bali, Sera Cremonini, Kedar Damle, Gergely Endr{\"o}di, Sean  Hartnoll, Elias Kiritsis,  Nilay Kundu, Gautam Mandal, Nilmani Mathur,  Kallol Sen, Aninda Sinha and Nandini Trivedi for discussions. We are especially grateful to Mohit Randeria 
for his  detailed discussions and insightful comments. 
SPT thanks the organisers of the ``Quantum Field theory, String Theory and Condensed Matter Physics'' held in Kolymbari, Crete,  1-7 September 2014,  and the organisers of the conference on 
``Perspectives and Challenges in Lattice Gauge Theory'' held in TIFR, Mumbai, 16-20th February, 2015.  
He  also thanks the Theory Division of CERN for hosting his sabbatical visit from June-December 2014, during which some of the research reported here was done. SPT acknowledges support from the DAE  and the    J. C. Bose Fellowship of  the Government of India. The work of S.J. is supported in part by the grant DOE 0000216873.
Most of all we thank the people of India for generously supporting research in String Theory. 

\appendix
\section{Numerical interpolation from the near horizon $AdS_3 \times R \times R$ to asymptotic $AdS_5$, }
\label{Numerics}
Our action consists of gravity, a massless dilaton $\phi$ and a cosmological constant $\Lambda$, in $5$ space time dimensions,
\be \label{actionint}
S_{bulk}=\frac{1}{2\kappa^2}\int d^{5}x\sqrt{-g}~\left(R+12 \Lambda-\frac{1}{2}(\partial\phi)^2\right).
\ee
Here $2\kappa^2=16\pi G$ is the gravitational coupling ($G$ is the Newton's Constant in 5 dimensions) and we set $\Lambda$=1.\\
It is easy to show that this system admits an $AdS_5$ solution with metric given by
\be
\label{ads5}
ds^2=\bigg[-u^2 dt^2 +{du^2 \over u^2} + u^2 (dx^2 + dy^2 + dz^2)\bigg],
\ee
and the dilaton is kept constant.

We now show that starting with the near horizon geometry given by eq.(\ref{nhz2scalar}), one can
 add a suitable perturbation which grows in the UV such that the solution matches asymptotically to  $AdS_5$ metric as provided in eq.\eqref{ads5}. \\
 This perturbation is given as follows-
\begin{align}
 \label{pertdef1}
\begin{split}
 g_{tt}(u) &=2 u^2  \left(1+\delta A(u)\right),\\
 g_{uu}(u) &=\frac{1 }{2 u^2  \left(1+\delta A(u)\right)}\\
 g_{xx}(u) &=2 u^2 \left(1+\delta A(u)\right), \\ 
 g_{yy}(u) &={\rho_{1}^2 \over 8} \left(1+\delta C(u)\right)\\
 g_{zz}(u) &={\rho_{2}^2 \over 8} \left(1+\delta D(u)\right)
 \end{split}
\end{align}
with 
\begin{align}
\delta A(u) =  a_1 \ u^{\nu},~ \delta C(u) =  c_2 \ u^{\nu},~  \delta D(u) =  c_3 \ u^{\nu} \\
a_1 =\frac{1}{5} (-5 + 2 \sqrt{5}) (c_{2} + c_{3}), ~ \nu=\sqrt{5}-1.
\end{align}
The numerical analysis is carried out using NDSolve in mathematica. For the case  $\rho_{1}=1$ ,  $\rho_{2}=1$ the suitably chosen values for $c_2 $ and $ c_3 $ are as follows
\be
\label{cval}
 c_2=85,~c_3=85.
\ee

By adjusting the coefficients $c_2,c_3$ to the above values one can ensure that the asymptotic behaviour of the metric eq.\eqref{pertdef1}
agrees with eq.(\ref{ads5}) at large u, say u$=$100000 ;

\begin{figure}
\begin{tabular}{cc}
\includegraphics[width=8cm]{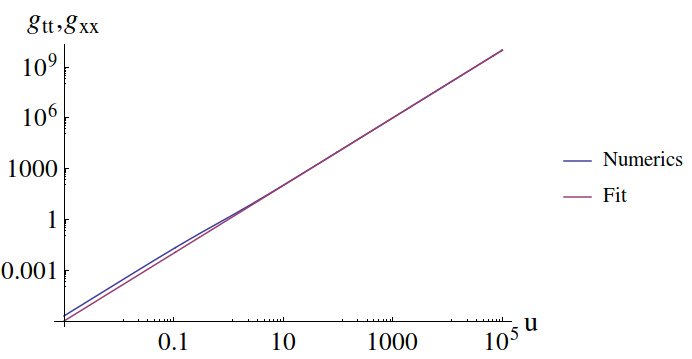}&\includegraphics[width=8cm]{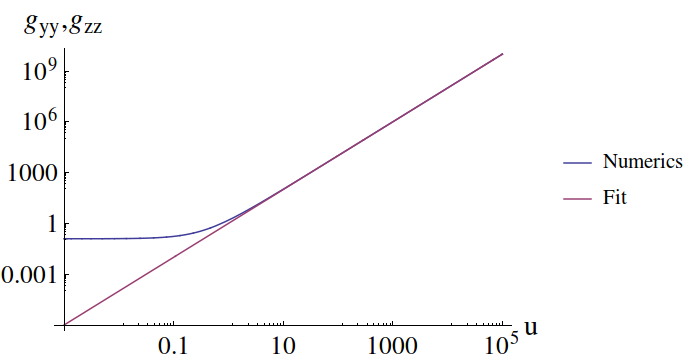}
	\end{tabular}
\caption{Log-log plot showing the numerical interpolation of near horizon $AdS_3 \times R \times R$ to asymptotic $AdS_5$ , with $\rho_{1}=1$ , $\rho_{2}=1$. 
}\label{num}
\end{figure}

The plots in Fig (\ref{num}) show the metric components as a function of $u$.
 These plots were obtained by numerical interpolation for the case $\rho_{1}=1$ ,  $\rho_{2}=1$ and $c_2=85 , c_3=85$ .

\section{Ratio of normalizable over non-normalizable mode near boundary}
\label{bdb}
 Here we check that asymptotically the canonical momentum  $\Pi$ goes to a constant independent of u . To see this , we consider the action 
% note that in the higher dimensional metric which is asymptotically $AdS_{5} $ , the metric perturbations behave as $ u^2 (1 + {C_{1}\over u^{4}})$ where  $C_{1}$ is constant. From the metric ansatz (\ref{5dMetric}) we find that $A_{\alpha} \sim  (1 + {C^{1}\over u^{4}})$ . From (\ref{pidef}) and (\ref{N}) , using the fact that we have asymtotic $AdS_{5}$ , we find that the canonical momentum  $\Pi$ goes to a constant ${2\over \kappa^{2}}C_{1}$ . Thus asysmptotically , the ratio of the normalizable to the non - normalizable mode behaves like ${2\over \kappa^{2}}C_{1}$.

% To be removed

\be 
S_{bulk}=\frac{1}{2\hat{\kappa}^2}\int d^{5}x\sqrt{-\hat {g}}~\left(\hat{R}+12 \Lambda\right).
\ee
we get the following solution for $AdS_{5}$ (setting $\Lambda$=1).
\be
\label{hads}
ds^2 =  \left(-u^{2}dt^{2} + \frac{du^{2}}{u^{2}}+u^{2}dx^{2} +  u^{2}dy^{2}+  u^{2}dz^{2}\right).
\ee
The metric perturbations go like  $ u^2 (1 + {C_{1}\over u^{4}})$ where  $C_{1}$ is constant.\\
Hence, using eq.(\ref{pidef}) and eq.(\ref{defN}) we find that
\be
\Pi(u)=-\frac{1}{2 \kappa^{2}} N(u) Z'= -\frac{1}{2 \kappa^{2}} \sqrt{-g}\frac{1}{g_{\rm{eff}}^2}g^{xx}g^{uu} \partial_{u} ({C_{1}\over u^{4}})=-\frac{1}{2 \kappa^{2}} \sqrt{-g} ~e^{3\psi}g^{xx}g^{uu} \partial_{u} ({C_{1}\over u^{4}}).
\ee
Plugging in the higher dimensional metric components from (\ref{hads})we get $\Pi(u)={2\over \kappa^{2}}C_{1}$ which is independent of u. Thus asymptotically, the ratio of the normalizable to the non - normalizable mode behaves like ${2\over \kappa^{2}}C_{1}$.

\section{Conductivity formula in terms of horizon quantities}
\label{b2hz}
In this appendix , we show the derivation of (\ref{exp33}) following \cite{Chakrabarti:2010xy}.
The electrical conductivity is defined in (\ref{defsig}) as
\begin{equation}
 \sigma(u,\omega) = \frac{\Pi(u,\omega)}{i \omega Z(u,w)}\big{|}_ {u\rightarrow \infty, \omega \rightarrow 0}.
\end{equation}

The real part can be written as
\begin{eqnarray}
 \Re(\sigma)  &=& \Re\Bigg(\frac{\Pi(u,\omega)}{i \omega Z(u,\omega)}\Bigg)\big{|}_ {u\rightarrow \infty, \omega \rightarrow 0}\nonumber
= \Re\Bigg(\frac{\Pi(u,\omega)Z(u,-\omega)}{i \omega Z(u,\omega)Z(u,-\omega)}\Bigg)\big{|}_ {u\rightarrow \infty, \omega \rightarrow 0}\nonumber\\
&=& \Im\Bigg(\frac{\Pi(u,\omega)Z(u,-\omega)}{ \omega Z^2(u)}\Bigg)\big{|}_ {u\rightarrow \infty, \omega \rightarrow 0}\nonumber
= {\Im\Bigg(\Pi(u,\omega)Z(u,-\omega)\Bigg)\over  \omega Z^2(u)}\big{|}_ {u\rightarrow \infty, \omega \rightarrow 0}.\\
\end{eqnarray}

Here we used the fact that $Z(u,\omega)\sim Z(u)$ is real to leading order when $\omega \rightarrow 0$.\\
We now proceed to show that\footnote{ $\Pi(u,\omega) ={\delta S \over \delta Z'(u,-\omega)}= -\frac{1}{2 \kappa^{2}} N(u) \frac{d}{du} Z(u,\omega) $ , hence $ \Im[\Pi(u,\omega)Z(u,-\omega)]$ behaves like a current .}
\begin{equation}
 \frac{d}{du}\Im[\Pi(u,\omega)Z(u,-\omega)] =0,
\end{equation}

This can be seen as follows
\begin{eqnarray}
\frac{d}{du}\Im \Big( N(u)\frac{d}{du}Z(u,\omega)Z(u,-\omega)\Big)\nonumber
=\Im \Big[ \frac{d}{du}\left(N(u)\frac{d}{du}Z(u,\omega)\right)Z(u,-\omega)\nonumber\\
+ N(u)\frac{d}{du}Z(u,\omega)\frac{d}{du}Z(u,-\omega)\Big].
\end{eqnarray}
Using (\ref{eqnmotion}), r.h.s of above equation reduces to
\begin{equation}
\Im\Big[- M(u)Z(u,\omega)Z(u,-\omega)+ N(u)\frac{d}{du}Z(u,\omega)\frac{d}{du}Z(u,-\omega)\Big]  ,
\end{equation}
which is equal to zero since the quantity in the bracket is real.
 Thus $\Im[\Pi(u,\omega)Z(u-\omega)]$ can be evaluated at the horizon i.e. at $u = u_{h}$.

Demanding regularity at the future horizon , we can approximate the behaviour of $Z(u,\omega)$ as follows 
\be
Z \sim e^{-i \omega (t+r_*)},
\ee
where $r_*$ is the tortoise coordinate,
\be
r_*=\int \sqrt{g_{uu}\over g_{tt}} \, du.
\ee

Using eq.(\ref{pidef})
and 
\begin{equation}
\lim_{u\rightarrow u_{h}}\frac{d}{du}Z(u,\omega)=-i \omega \lim_{u\to u_{h}}\sqrt{\frac{g_{uu}}{g_{tt}}} Z(u)+{\mathcal{O}}(\omega^{2}). 
\end{equation}
we get (in the limit $\omega \to 0$)

\begin{eqnarray}
\Re(\sigma) &=& \frac {1}{2\kappa^{2}} \Bigg(\sqrt{\frac{g_{uu}}{g_{tt}}}  N(u)\Bigg)_{u=u_{h}} \Bigg(\frac{Z(u_{h})}{Z(u\rightarrow \infty)}\Bigg)^{2} \nonumber\\
 &=& \frac {1}{2 \kappa^{2}}\Bigg(\sqrt{\frac{g_{uu}}{g_{tt}}}  ~\sqrt{-g}\frac{1}{g_{\rm{eff}}^2}g^{xx}g^{uu} \Bigg)_{u=u_{h}}\Bigg(\frac{Z(u_{h})}{Z(u\rightarrow \infty)}\Bigg)^{2}\nonumber\\
&=& \sigma_{H}~~\Bigg(\frac{Z(u_{h})}{Z(u\rightarrow \infty)}\Bigg)^{2} ,\label{sincleconduc11}
\end{eqnarray} 
where $\sigma_{H} $ is the conductivity evaluated at the horizon and its expression is given by,
\begin{equation}
\sigma_{H} = \frac {1}{2 \kappa^{2}g_{\rm{eff}}^2}\Big{|}_{u=u_{h}}.\end{equation}

where we used  isotropy along the spatial directions in the lower dimensional theory.

\bibliographystyle{jhepmod}
\bibliography{bibliography}

\providecommand{\href}[2]{#2}\begingroup\raggedright\begin{thebibliography}{10}

\bibitem{Policastro:2001yc}
G.~Policastro, D.~T. Son, and A.~O. Starinets, {\it {The Shear viscosity of
  strongly coupled N=4 supersymmetric Yang-Mills plasma}},  {\em
  Phys.Rev.Lett.} {\bf 87} (2001) p.~081601,
  [\href{http://xxx.lanl.gov/abs/hep-th/0104066}{{\tt hep-th/0104066}}].

\bibitem{Kovtun:2003wp}
P.~Kovtun, D.~T. Son, and A.~O. Starinets, {\it {Holography and hydrodynamics:
  Diffusion on stretched horizons}},  {\em JHEP} {\bf 0310} (2003) p.~064,
  [\href{http://xxx.lanl.gov/abs/hep-th/0309213}{{\tt hep-th/0309213}}].

\bibitem{Kovtun:2004de}
P.~Kovtun, D.~T. Son, and A.~O. Starinets, {\it {Viscosity in strongly
  interacting quantum field theories from black hole physics}},  {\em
  Phys.Rev.Lett.} {\bf 94} (2005) p.~111601,
  [\href{http://xxx.lanl.gov/abs/hep-th/0405231}{{\tt hep-th/0405231}}].

\bibitem{Kats:2007mq}
Y.~Kats and P.~Petrov, {\it {Effect of curvature squared corrections in AdS on
  the viscosity of the dual gauge theory}},  {\em JHEP} {\bf 0901} (2009)
  p.~044, [\href{http://xxx.lanl.gov/abs/0712.0743}{{\tt arXiv:0712.0743}}].

\bibitem{Buchel:2008vz}
A.~Buchel, R.~C. Myers, and A.~Sinha, {\it {Beyond eta/s = 1/4 pi}},  {\em
  JHEP} {\bf 0903} (2009) p.~084,
  [\href{http://xxx.lanl.gov/abs/0812.2521}{{\tt arXiv:0812.2521}}].

\bibitem{Sinha:2009ev}
A.~Sinha and R.~C. Myers, {\it {The Viscosity bound in string theory}},  {\em
  Nucl.Phys.} {\bf A830} (2009) pp.~295C--298C,
  [\href{http://xxx.lanl.gov/abs/0907.4798}{{\tt arXiv:0907.4798}}].

\bibitem{Cremonini:2011iq}
S.~Cremonini, {\it {The Shear Viscosity to Entropy Ratio: A Status Report}},
  {\em Mod.Phys.Lett.} {\bf B25} (2011) pp.~1867--1888,
  [\href{http://xxx.lanl.gov/abs/1108.0677}{{\tt arXiv:1108.0677}}].

\bibitem{Basu:2011tt}
P.~Basu and J.-H. Oh, {\it {Analytic Approaches to Anisotropic Holographic
  Superfluids}},  {\em JHEP} {\bf 1207} (2012) p.~106,
  [\href{http://xxx.lanl.gov/abs/1109.4592}{{\tt arXiv:1109.4592}}].

\bibitem{Bhattacharyya:2014wfa}
A.~Bhattacharyya and D.~Roychowdhury, {\it {Viscosity bound for anisotropic
  superfluids in higher derivative gravity}},  {\em JHEP} {\bf 1503} (2015)
  p.~063, [\href{http://xxx.lanl.gov/abs/1410.3222}{{\tt arXiv:1410.3222}}].

\bibitem{Brigante:2007nu}
M.~Brigante, H.~Liu, R.~C. Myers, S.~Shenker, and S.~Yaida, {\it {Viscosity
  Bound Violation in Higher Derivative Gravity}},  {\em Phys.Rev.} {\bf D77}
  (2008) p.~126006, [\href{http://xxx.lanl.gov/abs/0712.0805}{{\tt
  arXiv:0712.0805}}].

\bibitem{Brigante:2008gz}
M.~Brigante, H.~Liu, R.~C. Myers, S.~Shenker, and S.~Yaida, {\it {The Viscosity
  Bound and Causality Violation}},  {\em Phys.Rev.Lett.} {\bf 100} (2008)
  p.~191601, [\href{http://xxx.lanl.gov/abs/0802.3318}{{\tt arXiv:0802.3318}}].

\bibitem{Cohen:2007qr}
T.~D. Cohen, {\it {Is there a 'most perfect fluid' consistent with quantum
  field theory?}},  {\em Phys. Rev. Lett.} {\bf 99} (2007) p.~021602,
  [\href{http://xxx.lanl.gov/abs/hep-th/0702136}{{\tt hep-th/0702136}}].

\bibitem{Buchel:2010wf}
A.~Buchel and S.~Cremonini, {\it {Viscosity Bound and Causality in Superfluid
  Plasma}},  {\em JHEP} {\bf 10} (2010) p.~026,
  [\href{http://xxx.lanl.gov/abs/1007.2963}{{\tt arXiv:1007.2963}}].

\bibitem{Landsteiner:2007bd}
K.~Landsteiner and J.~Mas, {\it {The Shear viscosity of the non-commutative
  plasma}},  {\em JHEP} {\bf 0707} (2007) p.~088,
  [\href{http://xxx.lanl.gov/abs/0706.0411}{{\tt arXiv:0706.0411}}].

\bibitem{Azeyanagi:2009pr}
T.~Azeyanagi, W.~Li, and T.~Takayanagi, {\it {On String Theory Duals of
  Lifshitz-like Fixed Points}},  {\em JHEP} {\bf 0906} (2009) p.~084,
  [\href{http://xxx.lanl.gov/abs/0905.0688}{{\tt arXiv:0905.0688}}].

\bibitem{Natsuume:2010ky}
M.~Natsuume and M.~Ohta, {\it {The Shear viscosity of holographic
  superfluids}},  {\em Prog.Theor.Phys.} {\bf 124} (2010) pp.~931--951,
  [\href{http://xxx.lanl.gov/abs/1008.4142}{{\tt arXiv:1008.4142}}].

\bibitem{Erdmenger:2010xm}
J.~Erdmenger, P.~Kerner, and H.~Zeller, {\it {Non-universal shear viscosity
  from Einstein gravity}},  {\em Phys.Lett.} {\bf B699} (2011) pp.~301--304,
  [\href{http://xxx.lanl.gov/abs/1011.5912}{{\tt arXiv:1011.5912}}].

\bibitem{Erdmenger:2011tj}
J.~Erdmenger, P.~Kerner, and H.~Zeller, {\it {Transport in Anisotropic
  Superfluids: A Holographic Description}},  {\em JHEP} {\bf 1201} (2012)
  p.~059, [\href{http://xxx.lanl.gov/abs/1110.0007}{{\tt arXiv:1110.0007}}].

\bibitem{Mateos:2011ix}
D.~Mateos and D.~Trancanelli, {\it {The anisotropic N=4 super Yang-Mills plasma
  and its instabilities}},  {\em Phys.Rev.Lett.} {\bf 107} (2011) p.~101601,
  [\href{http://xxx.lanl.gov/abs/1105.3472}{{\tt arXiv:1105.3472}}].

\bibitem{Mateos:2011tv}
D.~Mateos and D.~Trancanelli, {\it {Thermodynamics and Instabilities of a
  Strongly Coupled Anisotropic Plasma}},  {\em JHEP} {\bf 1107} (2011) p.~054,
  [\href{http://xxx.lanl.gov/abs/1106.1637}{{\tt arXiv:1106.1637}}].

\bibitem{Cheng:2014sxa}
L.~Cheng, X.-H. Ge, and S.-J. Sin, {\it {Anisotropic plasma with a chemical
  potential and scheme-independent instabilities}},  {\em Phys.Lett.} {\bf
  B734} (2014) pp.~116--121, [\href{http://xxx.lanl.gov/abs/1404.1994}{{\tt
  arXiv:1404.1994}}].

\bibitem{Rebhan:2011vd}
A.~Rebhan and D.~Steineder, {\it {Violation of the Holographic Viscosity Bound
  in a Strongly Coupled Anisotropic Plasma}},  {\em Phys.Rev.Lett.} {\bf 108}
  (2012) p.~021601, [\href{http://xxx.lanl.gov/abs/1110.6825}{{\tt
  arXiv:1110.6825}}].

\bibitem{Polchinski:2012nh}
J.~Polchinski and E.~Silverstein, {\it {Large-density field theory, viscosity,
  and '$2k_F$' singularities from string duals}},  {\em Class.Quant.Grav.} {\bf
  29} (2012) p.~194008, [\href{http://xxx.lanl.gov/abs/1203.1015}{{\tt
  arXiv:1203.1015}}].

\bibitem{Giataganas:2012zy}
D.~Giataganas, {\it {Probing strongly coupled anisotropic plasma}},  {\em JHEP}
  {\bf 1207} (2012) p.~031, [\href{http://xxx.lanl.gov/abs/1202.4436}{{\tt
  arXiv:1202.4436}}].

\bibitem{Iizuka:2012wt}
N.~Iizuka and K.~Maeda, {\it {Study of Anisotropic Black Branes in
  Asymptotically anti-de Sitter}},  {\em JHEP} {\bf 1207} (2012) p.~129,
  [\href{http://xxx.lanl.gov/abs/1204.3008}{{\tt arXiv:1204.3008}}].

\bibitem{Mamo:2012sy}
K.~A. Mamo, {\it {Holographic RG flow of the shear viscosity to entropy density
  ratio in strongly coupled anisotropic plasma}},  {\em JHEP} {\bf 1210} (2012)
  p.~070, [\href{http://xxx.lanl.gov/abs/1205.1797}{{\tt arXiv:1205.1797}}].

\bibitem{Jain:2014vka}
S.~Jain, N.~Kundu, K.~Sen, A.~Sinha, and S.~P. Trivedi, {\it {A Strongly
  Coupled Anisotropic Fluid From Dilaton Driven Holography}},  {\em JHEP} {\bf
  1501} (2015) p.~005, [\href{http://xxx.lanl.gov/abs/1406.4874}{{\tt
  arXiv:1406.4874}}].

\bibitem{Critelli:2014kra}
R.~Critelli, S.~Finazzo, M.~Zaniboni, and J.~Noronha, {\it {Anisotropic shear
  viscosity of a strongly coupled non-Abelian plasma from magnetic branes}},
  {\em Phys.Rev.} {\bf D90} (2014), no.~6 p.~066006,
  [\href{http://xxx.lanl.gov/abs/1406.6019}{{\tt arXiv:1406.6019}}].

\bibitem{Ge:2014aza}
X.-H. Ge, Y.~Ling, C.~Niu, and S.-J. Sin, {\it {Holographic transports and
  stability in anisotropic linear axion model}},
  \href{http://xxx.lanl.gov/abs/1412.8346}{{\tt arXiv:1412.8346}}.

\bibitem{Witten:1998zw}
E.~Witten, {\it {Anti-de Sitter space, thermal phase transition, and
  confinement in gauge theories}},  {\em Adv.Theor.Math.Phys.} {\bf 2} (1998)
  pp.~505--532, [\href{http://xxx.lanl.gov/abs/hep-th/9803131}{{\tt
  hep-th/9803131}}].

\bibitem{Klebanov:2000hb}
I.~R. Klebanov and M.~J. Strassler, {\it {Supergravity and a confining gauge
  theory: Duality cascades and chi SB resolution of naked singularities}},
  {\em JHEP} {\bf 0008} (2000) p.~052,
  [\href{http://xxx.lanl.gov/abs/hep-th/0007191}{{\tt hep-th/0007191}}].

\bibitem{Iqbal:2008by}
N.~Iqbal and H.~Liu, {\it {Universality of the hydrodynamic limit in AdS/CFT
  and the membrane paradigm}},  {\em Phys.Rev.} {\bf D79} (2009) p.~025023,
  [\href{http://xxx.lanl.gov/abs/0809.3808}{{\tt arXiv:0809.3808}}].

\bibitem{D'Hoker:2009mm}
E.~D'Hoker and P.~Kraus, {\it {Magnetic Brane Solutions in AdS}},  {\em JHEP}
  {\bf 0910} (2009) p.~088, [\href{http://xxx.lanl.gov/abs/0908.3875}{{\tt
  arXiv:0908.3875}}].

\bibitem{Myers:2009ij}
R.~C. Myers, M.~F. Paulos, and A.~Sinha, {\it {Holographic Hydrodynamics with a
  Chemical Potential}},  {\em JHEP} {\bf 0906} (2009) p.~006,
  [\href{http://xxx.lanl.gov/abs/0903.2834}{{\tt arXiv:0903.2834}}].

\bibitem{Chakrabarti:2010xy}
S.~K. Chakrabarti, S.~Chakrabortty, and S.~Jain, {\it {Proof of universality of
  electrical conductivity at finite chemical potential}},  {\em JHEP} {\bf
  1102} (2011) p.~073, [\href{http://xxx.lanl.gov/abs/1011.3499}{{\tt
  arXiv:1011.3499}}].

\bibitem{Maharana:1992my}
J.~Maharana and J.~H. Schwarz, {\it {Noncompact symmetries in string theory}},
  {\em Nucl.Phys.} {\bf B390} (1993) pp.~3--32,
  [\href{http://xxx.lanl.gov/abs/hep-th/9207016}{{\tt hep-th/9207016}}].

\bibitem{Landau1987Fluid}
L.~D. Landau and E.~M. Lifshitz, {\it {\em Fluid Mechanics, Second Edition:
  Volume 6 (Course of Theoretical Physics)}},  {\em Butterworth-Heinemann,
  2~ed.} (Jan., 1987).

\bibitem{REI65}
F.~Reif, {\it {\em Fundamentals of Statistical and Thermal Physics}},  {\em
  McGraw Hill, Tokyo} (1965).

\bibitem{Bocquet:1995je}
M.~Bocquet, S.~Bonazzola, E.~Gourgoulhon, and J.~Novak, {\it {Rotating neutron
  star models with magnetic field}},  {\em Astron.Astrophys.} {\bf 301} (1995)
  p.~757, [\href{http://xxx.lanl.gov/abs/gr-qc/9503044}{{\tt gr-qc/9503044}}].

\bibitem{Harding:2006qn}
A.~K. Harding and D.~Lai, {\it {Physics of Strongly Magnetized Neutron Stars}},
   {\em Rept.Prog.Phys.} {\bf 69} (2006) p.~2631,
  [\href{http://xxx.lanl.gov/abs/astro-ph/0606674}{{\tt astro-ph/0606674}}].

\bibitem{Endrodi:2014lja}
G.~Endrödi, {\it {Magnetic structure of isospin-asymmetric QCD matter in
  neutron stars}},  {\em Phys.Rev.} {\bf D90} (2014), no.~9 p.~094501,
  [\href{http://xxx.lanl.gov/abs/1407.1216}{{\tt arXiv:1407.1216}}].

\bibitem{Kharzeev:2007jp}
D.~E. Kharzeev, L.~D. McLerran, and H.~J. Warringa, {\it {The Effects of
  topological charge change in heavy ion collisions: 'Event by event P and CP
  violation'}},  {\em Nucl.Phys.} {\bf A803} (2008) pp.~227--253,
  [\href{http://xxx.lanl.gov/abs/0711.0950}{{\tt arXiv:0711.0950}}].

\bibitem{Fukushima:2008xe}
K.~Fukushima, D.~E. Kharzeev, and H.~J. Warringa, {\it {The Chiral Magnetic
  Effect}},  {\em Phys.Rev.} {\bf D78} (2008) p.~074033,
  [\href{http://xxx.lanl.gov/abs/0808.3382}{{\tt arXiv:0808.3382}}].

\bibitem{Skokov:2009qp}
V.~Skokov, A.~Y. Illarionov, and V.~Toneev, {\it {Estimate of the magnetic
  field strength in heavy-ion collisions}},  {\em Int.J.Mod.Phys.} {\bf A24}
  (2009) pp.~5925--5932, [\href{http://xxx.lanl.gov/abs/0907.1396}{{\tt
  arXiv:0907.1396}}].

\bibitem{2014Sci...344..722B}
A.~B. {Bardon}, S.~{Beattie}, C.~{Luciuk}, W.~{Cairncross}, D.~{Fine}, N.~S.
  {Cheng}, G.~J.~A. {Edge}, E.~{Taylor}, S.~{Zhang}, S.~{Trotzky}, and J.~H.
  {Thywissen}, {\it {Transverse Demagnetization Dynamics of a Unitary Fermi
  Gas}},  {\em Science} {\bf 344} (May, 2014) pp.~722--724,
  [\href{http://xxx.lanl.gov/abs/1310.5140}{{\tt arXiv:1310.5140}}].

\bibitem{2013NatPh...9..405K}
M.~{Koschorreck}, D.~{Pertot}, E.~{Vogt}, and M.~{K{\"o}hl}, {\it {Universal
  spin dynamics in two-dimensional Fermi gases}},  {\em Nature Physics} {\bf 9}
  (July, 2013) pp.~405--409, [\href{http://xxx.lanl.gov/abs/1304.4980}{{\tt
  arXiv:1304.4980}}].

\bibitem{Domokos:2007kt}
S.~K. Domokos and J.~A. Harvey, {\it {Baryon number-induced Chern-Simons
  couplings of vector and axial-vector mesons in holographic QCD}},  {\em
  Phys.Rev.Lett.} {\bf 99} (2007) p.~141602,
  [\href{http://xxx.lanl.gov/abs/0704.1604}{{\tt arXiv:0704.1604}}].

\bibitem{Rozali:2007rx}
M.~Rozali, H.-H. Shieh, M.~Van~Raamsdonk, and J.~Wu, {\it {Cold Nuclear Matter
  In Holographic QCD}},  {\em JHEP} {\bf 0801} (2008) p.~053,
  [\href{http://xxx.lanl.gov/abs/0708.1322}{{\tt arXiv:0708.1322}}].

\bibitem{Nakamura:2009tf}
S.~Nakamura, H.~Ooguri, and C.-S. Park, {\it {Gravity Dual of Spatially
  Modulated Phase}},  {\em Phys.Rev.} {\bf D81} (2010) p.~044018,
  [\href{http://xxx.lanl.gov/abs/0911.0679}{{\tt arXiv:0911.0679}}].

\bibitem{Ooguri:2010kt}
H.~Ooguri and C.-S. Park, {\it {Holographic End-Point of Spatially Modulated
  Phase Transition}},  {\em Phys.Rev.} {\bf D82} (2010) p.~126001,
  [\href{http://xxx.lanl.gov/abs/1007.3737}{{\tt arXiv:1007.3737}}].

\bibitem{Ooguri:2010xs}
H.~Ooguri and C.-S. Park, {\it {Spatially Modulated Phase in Holographic
  Quark-Gluon Plasma}},  {\em Phys.Rev.Lett.} {\bf 106} (2011) p.~061601,
  [\href{http://xxx.lanl.gov/abs/1011.4144}{{\tt arXiv:1011.4144}}].

\bibitem{Bergman:2011rf}
O.~Bergman, N.~Jokela, G.~Lifschytz, and M.~Lippert, {\it {Striped instability
  of a holographic Fermi-like liquid}},  {\em JHEP} {\bf 1110} (2011) p.~034,
  [\href{http://xxx.lanl.gov/abs/1106.3883}{{\tt arXiv:1106.3883}}].

\bibitem{Donos:2011qt}
A.~Donos, J.~P. Gauntlett, and C.~Pantelidou, {\it {Spatially modulated
  instabilities of magnetic black branes}},  {\em JHEP} {\bf 1201} (2012)
  p.~061, [\href{http://xxx.lanl.gov/abs/1109.0471}{{\tt arXiv:1109.0471}}].

\bibitem{Donos:2011ff}
A.~Donos and J.~P. Gauntlett, {\it {Holographic helical superconductors}},
  {\em JHEP} {\bf 1112} (2011) p.~091,
  [\href{http://xxx.lanl.gov/abs/1109.3866}{{\tt arXiv:1109.3866}}].

\bibitem{Donos:2012gg}
A.~Donos and J.~P. Gauntlett, {\it {Helical superconducting black holes}},
  {\em Phys.Rev.Lett.} {\bf 108} (2012) p.~211601,
  [\href{http://xxx.lanl.gov/abs/1203.0533}{{\tt arXiv:1203.0533}}].

\bibitem{Donos:2012wi}
A.~Donos and J.~P. Gauntlett, {\it {Black holes dual to helical current
  phases}},  {\em Phys.Rev.} {\bf D86} (2012) p.~064010,
  [\href{http://xxx.lanl.gov/abs/1204.1734}{{\tt arXiv:1204.1734}}].

\bibitem{2012JHEP...07..193I}
N.~{Iizuka}, S.~{Kachru}, N.~{Kundu}, P.~{Narayan}, N.~{Sircar}, and S.~P.
  {Trivedi}, {\it {Bianchi attractors: a classification of extremal black brane
  geometries}},  {\em Journal of High Energy Physics} {\bf 7} (July, 2012)
  p.~193, [\href{http://xxx.lanl.gov/abs/1201.4861}{{\tt arXiv:1201.4861}}].

\bibitem{Iizuka:2012pn}
N.~Iizuka, S.~Kachru, N.~Kundu, P.~Narayan, N.~Sircar, {\em et.~al.}, {\it
  {Extremal Horizons with Reduced Symmetry: Hyperscaling Violation, Stripes,
  and a Classification for the Homogeneous Case}},  {\em JHEP} {\bf 1303}
  (2013) p.~126, [\href{http://xxx.lanl.gov/abs/1212.1948}{{\tt
  arXiv:1212.1948}}].

\bibitem{Ovdat:2014ipa}
O.~Ovdat and A.~Yarom, {\it {A modulated shear to entropy ratio}},  {\em JHEP}
  {\bf 1411} (2014) p.~019, [\href{http://xxx.lanl.gov/abs/1407.6372}{{\tt
  arXiv:1407.6372}}].

\end{thebibliography}\endgroup
\end{document}